\documentclass[11pt]{article}
\usepackage{cite}
\usepackage{amsmath,amsfonts,amssymb}
\usepackage[small,bf,hang]{caption}
\usepackage{slashed}
\usepackage[latin5]{inputenc}

\def\hybrid{
        \topmargin -20pt
        \oddsidemargin 0pt
        \headheight 0pt \headsep 0pt
        \textwidth 6.25in 
        \textheight 9.5in 
        \marginparwidth .875in
        \parskip 5pt plus 1pt \jot = 1.5ex}

\hybrid

 \csname
@addtoreset\endcsname{equation}{section}

\def\moth{\mathsurround=0pt}
\newdimen\zo \zo=0pt

\def\tick{\leaders\hrule height 0.5ex depth 0pt \hskip 0.5pt}
\def\upboxfill{$\moth \setbox\zo\hbox{\tick}%
  \hskip 3pt\hbox to 0pt{$\tick$\hss}\hrulefill \hbox to 7.5pt{$\tick$\hss}$}

\def\dtick{\leaders\hrule height .34pt depth 0.5ex \hskip 0.5pt}
\def\downboxfill{$\moth \setbox\zo\hbox{\dtick}%
  \hskip 2pt\hbox to 0pt{$\dtick$\hss}\hrulefill \hbox to 2pt{$\dtick$\hss}$}

\def\ciftS{\mathbb{S}}

\def\cH{{\cal H}}
\def\cK{{\cal K}}
\def\cR{{\cal R}}

\def\del{\partial}

\def\be{\begin{equation}}
\def\ee{\end{equation}}
\def\bea{\begin{eqnarray}}
\def\eea{\end{eqnarray}}
\def\ba{\begin{array}}
\def\ea{\end{array}}

\begin{document}

\begin{titlepage}
\rightline{} \rightline\today
\begin{center}
\vskip 2.5cm {\Large \bf {Non-Abelian T-duality as a Transformation in Double Field Theory}}\\
\vskip 2.5cm {\large {Aybike \c{C}atal-\"{O}zer}} \vskip 1cm
{\it {Department of Mathematics,}}\\
{\it {\.{I}stanbul Technical University,}}\\
{\it {Maslak 34469,
\.{I}stanbul, Turkey}}\\
ozerayb@itu.edu.tr \vskip 0.5cm

\vskip 1cm {\bf Abstract}
\end{center}

\vskip 0.5cm

\noindent
\begin{narrower}

\noindent Non-Abelian T-duality (NATD) is a solution generating
transformation for supergravity backgrounds with non-Abelian
isometries. We show that NATD  can be described as a coordinate
dependent O(d,d) transformation, where the dependence on the
coordinates is determined by the structure constants of the Lie
algebra associated with the isometry group.  Besides making
calculations significantly easier, this approach gives a natural
embedding of NATD  in Double Field Theory (DFT), a framework which
provides an O(d,d) covariant formulation for effective string
actions. As a result of this embedding, it becomes easy to prove
that the NATD transformed backgrounds solve supergravity
equations, when the isometry algebra is unimodular. If the
isometry algebra is non-unimodular, the generalized dilaton field
is forced to have a linear dependence on the dual coordinates,
which implies that the resulting background solves generalized
supergravity equations.

\end{narrower}

\vspace{4cm}

\end{titlepage}

\newpage

\tableofcontents

\newpage

\section{Introduction}\label{introduction}

Non-Abelian T-duality (NATD) is a generalization of T-duality for
strings on backgrounds with non-Abelian isometries
\cite{delaOssa:1992vci}-\cite{Alvarez:1994np}. Although it is not
as well established as T-duality is as a string duality symmetry,
it works well as a solution generating transformation for
supergravity. The rules for the transformation of the fields in
the NS-NS sector, namely the metric, the B-field and the dilaton
field has been known for a long time. Recently, NATD has gained a
new interest, as the rules for the transformation of the fields in
the RR sector of Type II strings has also been found
\cite{Sfetsos:2010uq}. This has been applied to many supergravity
backgrounds by various groups, especially to backgrounds that are
relevant for AdS-CFT correspondence, see for example
\cite{Lozano:2011kb}-\cite{Itsios:2017cew}.

Recently, a compact formula for the transformation of the
supergravity fields for a generic Green-Schwarz string with
isometry $G$  has been obtained in \cite{Borsato:2018idb}, where
they also showed that the sigma model after NATD has kappa
symmetry. This  means that the resulting background is a solution
of the generalized supergravity equations (GSE), which have
recently been introduced in \cite{Arutyunov:2015mqj} as a
generalization of supergravity equations, see also
\cite{Wulff:2016tju}. To be more precise, when the isometry group
$G$ is unimodular, the dualized sigma model is Weyl invariant and
the target space is  a solution of standard Type II supergravity
equations. If $G$ is non-unimodular so that the structure
constants of the Lie algebra of $G$ is not traceless, the trace
components give rise to a deformation of the equations to be
satisfied by the target space fields to GSE\footnote{The fact that
the NATD background fails to satisfy standard supergravity
equations when $G$ is non-unimodular was first noted in
\cite{veneziano1} and the generalized equations  appeared first in
\cite{veneziano2}. For a detailed account, see
\cite{Hong:2018tlp}.}.

The purpose of this paper is to  describe the NATD transformation
rules obtained in \cite{Borsato:2018idb} as a coordinate dependent
$O(10,10)$ transformation\footnote{The $O(10,10)$ matrix
associated with the NATD transformation is obtained by embedding
an $O(d,d)$ matrix in $O(10,10)$, where $d$ is the dimension of
the isometry group. Hence, the only non-trivial action is on the
isometry directions. For this reason, we will sometimes refer to
this action as an $O(d,d)$ transformation.}. In Abelian T-duality
with $d$ commuting isometries, the transformation rules for the
supergravity fields in the NS-NS sector can be neatly described
through the action of a constant $O(d,d)$ matrix embedded in
$O(10,10)$ \cite{Giveon:1994fu}. The RR fields are then packaged
in a differential form, which can be a regarded as a spinor field
that transforms under $Spin(d,d)$. If the fields in the NS-NS
sector transform under $T \in O(d,d)$, then the spinor field that
encodes the RR fields transform under $S_T \in Spin(d,d)$, which
is the element that projects onto $T$ under the double covering
homomorphism $\rho$ between $O(d,d)$ and $Spin(d,d)$, that is,
$\rho(S) = T$ \cite{Fukuma}. In a similar fashion, we show in this
paper that the NATD transformation of the supergravity fields in
the NS-NS sector can be described through the action of an
$O(10,10)$ matrix (presented in (\ref{NATDmatrix})), this time not
constant but with an explicit dependence on the coordinates of the
dual theory. The dependence on the coordinates is determined by
the structure constants of the Lie algebra associated with the
isometry group. The transformation of the RR fields is then
automatically determined by the corresponding $Spin(d,d)$ matrix,
as in Abelian T-duality. We would like to note that we had already
presented the NATD matrix we give in this paper at a workshop at
APCTP, Pohang \cite{koreatalk}. Very recently, a paper has
appeared which also views NATD as an $O(d,d)$ transformation
\cite{Sakatani:2019jgu}. See also \cite{Bugden:2019vlj}, which has
a similar approach to NATD.

Besides making calculations significantly easier, our approach
makes it possible to view NATD as a solution generating
transformation in Double Field Theory (DFT), a framework which
provides an $O(d,d)$ covariant formulation for effective string
actions \cite{Tseytlin1}-\cite{dftRR} by introducing dual, winding
type coordinates. In its current formulation, DFT is a consistent
field theory only when a certain constraint, called the strong
constraint is satisfied. If the DFT fields have no dependence on
the winding type coordinates, the strong constraint is satisfied
trivially and the fields are said to be in the supergravity frame.
In such a case, DFT of Type II strings constructed in
\cite{HullZ4, dftRR} reduces to Type II supergravity in the
democratic formulation. When the duality group is unimodular, the
NATD fields always belong to the supergravity frame, and hence our
method provides a simple proof of the fact that the transformed
fields solve supergravity equations, when the isometry algebra is
unimodular. If the isometry algebra is non-unimodular, we show
that the generalized dilaton field of DFT is forced to have a
linear dependence on the winding type coordinates. In such a
coordinate frame, DFT equations are known to reduce to generalized
supergravity equations \cite{Yoshida1, Yoshida2}. This then
implies that the resulting NATD fields should solve GSE,
consistent with what has been found in the literature so far. Let
us make a remark at this point. The NATD rules for the NS-NS
sector and the fact that it is a solution generating
transformation for supergravity has been known since early 90's,
as has been mentioned before. These rules were obtained directly
from the sigma model by applying the Buscher procedure. On the
other hand, the rules for the fields in the RR sector has been
figured out only recently in \cite{Sfetsos:2010uq}, by "guessing"
them from how the Buscher rules extend to the RR sector in Abelian
T-duality (and initially only for the Principal Chiral Model).
This approach does not provide a proof of why the transformed
fields should constitute a proper supergravity background, so it
had to be checked example by example that  the NATD fields indeed
solved (generalized) supergravity equations. Embedding NATD in DFT
as we do  here provides a proof that this should always be the
case. Our approach in this paper is quite different from that of
\cite{Borsato:2018idb}, where they also prove that the dualized
fields satisfy the GSE by checking the kappa symmetry of the
transformed Green-Schwarz sigma model. Here, we consider the
transformation of the fields within DFT and directly check that
the dual fields satisfy the field equations of DFT in an
appropriate frame, where they reduce to
 (generalized) supergravity equations. NATD
  has been studied in the context of DFT also in the
papers  \cite{Lust:2018jsx}, \cite{Demulder:2018lmj}.
 There is a generalized notion of T-duality, called the Poisson Lie T-Duality \cite{klimcik1,klimcik2},
  which does not require the symmetry group $G$ to act by isometries. It includes Abelian T-duality and NATD as special cases. Poisson Lie T-duality has been studied
 in the context of DFT in \cite{Hassler:2017yza}\footnote{To be more precise,  \cite{Hassler:2017yza} studies Poisson Lie T-duality   in the framework of
 DFT on group manifolds (usually called DFT$_{{\rm WZW}}$), which is a different theory from the standard DFT. DFT on group manifolds was constructed in the papers
 \cite{hassler1,hassler2,hassler3}.}  and very recently in \cite{Sakatani:2019jgu}.

As the $O(10,10)$ matrix that produces the NATD fields is not
constant, it is not immediately clear that it generates  a
solution generating transformation for DFT. To show that this is
indeed the case, we find it useful to utilize the framework of
Gauged Double Field Theory (GDFT), which is obtained by a duality
twisted (Scherk-Schwarz) reduction \cite{SS} of DFT
\cite{Geissbuhler}-\cite{Aybike}. GDFT is a deformation of DFT,
determined by the \emph{fluxes} associated with the twist matrix
that define the duality twisted reduction anzats. Our strategy is
as follows: We start with a solution of Type II supergravity.
Since Type II supergravity can be embedded in DFT, one can
construct corresponding DFT fields which constitute a solution for
DFT in the supergravity frame. If the space-time metric has an
isometry symmetry $G$, which is also respected by the B-field and
the RR fluxes (not necessarily the gauge potentials), we can
extract DFT fields out of the original ones, which satisfy the
field equations of GDFT determined by the \emph{geometric flux}
associated with the isometry group $G$. We call these fields
\emph{untwisted} DFT fields.  The DFT fields corresponding to the
NATD background are obtained by acting on these untwisted fields
with the $O(10,10)$ NATD matrix we present here. We will show that
these dual fields also satisfy the field equations of DFT by using
the following three key facts, which we will prove in the body of
the paper: \emph{i)} Field equations of GDFT are $O(d,d)$ (or
$Spin(d,d)$) covariant, provided we also allow fluxes transform as
generalized tensors. \emph{ii)} By using fact \emph{(i)} above, we
show that a set of duality twisted DFT fields, which we
generically write as $\phi(x, Y) = U(Y) . \phi(x)$\footnote{Here,
the action of $U(Y) \in O(d,d)$ is determined by how $\phi$
transforms under $O(d,d)$ or more generally under $Spin(d,d)$ if
$\phi$ is a spinor field in DFT.} satisfy the field equations of
DFT if and only if the untwisted fields $\phi(x)$ satisfy the
field equations of GDFT determined by the fluxes associated by the
twist matrix $U$. This immediately implies the following: a set of
fields $\tilde{\phi}(x, Z) = \tilde{U}(Z). \phi(x)$, where the
twist matrix $\tilde{U}(Z)$ produces the same fluxes as $U(Y)$,
will satisfy the field equations of DFT if and only if the fields
$\phi(x, Y) = U(Y) . \phi(x)$ satisfy the field equations DFT.
\emph{iii)} The fluxes associated with the isometry group $G$ and
the NATD matrix $T$ are exactly the same. Facts \emph{ii)} and
\emph{iii)} together prove that the NATD fields indeed form a
solution of DFT, as claimed.

The structure of the paper is as follows: In the next section, we
give a brief review of the $O(d,d)$ structure of Abelian
T-duality, first in the NS-NS sector and then in RR sector in
subsection (\ref{RRsector}). This enables us to identify the
coordinate dependent $O(d,d)$ matrix that generates the NATD
background in section (\ref{NATD}). Also in this section (in
subsection (\ref{example})), we demonstrate how the NATD of the
background $AdS_3 \times S_3 \times T^4$ studied in
\cite{Sfetsos:2010uq} can be obtained by the action of the NATD
matrix we have identified. Then in section (\ref{DFT}), we study
the embedding of NATD in DFT. The three key facts that we listed
in the previous paragraph are proved in this section. The
distinction between the unimodular and non-unimodular cases is
also discussed here. We finish the paper with discussions and
outlook in section (\ref{conclusion}).

\textbf{ Note Added:} While we were about to finalize the writing
of this manuscript, the paper \cite{Sakatani:2019jgu} appeared on
the arXiv, parts of which  overlap with the work we present here.

\section{The Action of $O(d,d)$ on curved string
backgrounds}\label{Odd}

In this section, we review how Abelian T-duality can be described
as an $O(d,d)$ transformation, first for the NS-NS sector and then
for the RR sector. For the RR sector, the duality group should be
lifted to $Spin(d,d)$. We closely follow \cite{Giveon:1994fu} in
section (\ref{NSsector}) and \cite{Fukuma} in section
(\ref{RRsector}).

\subsection{Transformation of the fields in the NS-NS sector}\label{NSsector}

Let $g$ and $B$ be the metric and the Kalb-Ramond 2-form field
that describes a $D$ dimensional supergravity background, with d
commuting isometries. The string living on this background
exhibits an $O(d,d,Z)$ T-duality symmetry. Accordingly, there is
an $O(d,d,R)$ action, which acts as a solution generating symmetry
in the low energy limit. Let us define the $D \times D$ background
matrix \be \label{background} Q(G, B) =G + B. \ee Due to the
presence of d commuting isometries, it is possible to choose
adopted coordinates $X = (x^{I}, x^{m}), I= 1, \cdots, d $ such
that the background matrix does not depend on the $d$ coordinates
$x^{I}$. Let us decompose the background
matrix with respect to this choice of coordinates as \be Q = \left(\begin{array}{cc} Q_{IJ} & Q_{Im} \\
                           Q_{mI} & Q_{mn}
                           \end{array}\right)=
\left(\begin{array}{cc} E & F^2 \\
                           F^1 & F \end{array}\right). \ee

Let $T$ be a matrix in $O(d,d,R)$. Then \be T =
\left(\begin{array}{cc} a & b \\
                        c & d \end{array}\right), \ \ \ a^t c +
                        c^t a = 0, \ \ b^t d + d^t b = 0, \ \ a^t
                        d + c^t b = I. \ee

This can be embedded in $O(D,D,R)$ as follows \be
 \hat{T} =
\left(\begin{array}{cc} \label{matrixT} \hat{a} & \hat{b} \\
                        \hat{c} & \hat{d} \end{array}\right), \ee
                        where $\hat{a}, \hat{b}, \hat{c}, \hat{d}$
                        are $D \times D$ matrices defined below:

\be \label{embedding} \hat{a} = \left(\begin{array}{cc} a & 0 \\
                                      0 & I \end{array}\right), \
                                      \ \hat{b} = \left(\begin{array}{cc} b & 0 \\
                                      0 & 0 \end{array}\right), \
                                      \ \hat{c} = \left(\begin{array}{cc} c & 0 \\
                                      0 & 0 \end{array}\right), \
                                      \ \hat{d} = \left(\begin{array}{cc} d & 0 \\
                                      0 & I \end{array}\right).
                                      \ee

Let \be \label{finalQ} Q'(G', B') = \hat{T} . \ Q(G, B) \equiv
(\hat{a} Q + \hat{b})(\hat{c} Q + \hat{d})^{-1} \ee be the new
background matrix obtained by the above action of $O(D,D,R)$ on
$Q$. Then, it is well known that the transformed metric and the
transformed B-field obtained from \be G' = \frac{Q' + Q'^t}{2}, \
\ B' = \frac{Q' - Q'^t}{2} \ee define (along with the transformed
dilaton field we will discuss below, see (\ref{dilaton1}))  valid
supergravity backgrounds. That is, the
  $O(D,D,R)$ transformation defined above acts as a solution
  generating transformation.

  For completeness, let us write the final form of the transformed
  background matrix $Q'$:

  \be \label{genelQ} Q' = \left(\begin{array}{cc} E' & (a - E' c) F^2 \\
                                   F^1(cE + d)^{-1} & F - F^1(cE +
                                   d)^{-1} c F^2 \end{array} \right), \ee
                                   where \be \label{genelE} E' = T . E = (a E + b)(c E +
                                   d)^{-1}. \ee

For the resulting background to be a valid supergravity solution,
the dilaton field $\phi$ should also transform under $O(D,D,R)$ in
the following way: \be \label{dilaton1} e^{-2 \phi'} = e^{-2 \phi}
\sqrt{\frac{{\rm det }G}{{\rm det }G'}}  \ee For later reference,
we define the following field $d$, which is invariant under
$O(D,D)$ transformations: \be \label{gendilaton} d = \phi -
\frac{1}{4} \ln{{\rm det }G}. \ee  Note that this gives  \be
\label{gendilaton2} e^{-2d} = \sqrt{{\rm det} G } \ e^{-2 \phi}.
\ee It is easily checked from (\ref{dilaton1}) that $e^{-2d'} =
e^{-2d}$ under $O(D,D)$. The field $d$ will play an important role
in DFT as the generalized dilaton field, as we will discuss in
section  \ref{DFTreview}.


\subsection{Transformation of the fields in the RR sector}\label{RRsector}

Let us now discuss how the p-form fields in the RR sector of Type
II supergravity theory transform under the action of $O(d,d,R)$
described in the section above. For this discussion we closely
follow \cite{Fukuma}, see also \cite{CatalOzer:2005mr}.

In the democratic formulation of Type II supergravity
\cite{Bergshoeff}, the 0,2 and 4-form fields in Type IIB and 1 and
3 form fields in Type IIA are combined with their Hodge duals to
form sections of the exterior bundle $\bigwedge^{{\rm even}} T^*M$
for the first case and of $\bigwedge^{{\rm odd}} T^*M$ for the
latter, where $M$ is the  space-time manifold. It is well known
that these bundles carry the chiral spinor representations of
$Pin(d,d)$, which is the double covering group of $O(d,d)$. The
transformation of the RR fields under T-duality is determined by
this action of $Pin(d,d)$ on the RR fields, viewed as a section of
the exterior bundle. More precisely, if the T-duality
transformation in the NS-NS sector is realized by the $O(d,d)$
matrix $T$, then the $Pin(d,d)$ element acting on the spinor field
that packages the modified RR gauge potentials is $S$, where
$\rho(S) = T$.\footnote{See the papers \cite{dftRR} and
\cite{Aybike} for an overview of $O(d,d)$ and its double covering
group $Pin(d,d)$. The method to find the $Pin(d,d)$ matrix
corresponding to a given $O(d,d)$ matrix is also explained in
these papers.} Here, $\rho$ is the double covering map \be
\label{coveringmap} \rho : Pin(d,d) \rightarrow O(d,d). \ee Then,
if $\chi$ is the spinor field that packages the modified RR fields
we have \be \label{transfchi}
 \chi \rightarrow \chi' = S \ \chi \ee Let us now discuss the
 transformation of the field strength $\slashed{\del}\chi$ under
 $Pin(d,d)$. This is important since RR fluxes are defined as \be
 \label{ek1}
 F = e^{-B} \slashed{\del}\chi. \ee As was discussed in \cite{Fukuma}, the
 transformation (\ref{transfchi}) does not imply
 $\slashed{\del}\chi \rightarrow S \slashed{\del}\chi$. However,
 when one doubles the space-time coordinates as in Double Field
 Theory (DFT), which we will discuss in more detail in section (\ref{DFTreview}), $\slashed{\del}\chi$
 also transforms as a vector under $Pin(d,d)$ as $\slashed{\del}\chi \rightarrow S
 \slashed{\del}\chi$. In DFT,  the usual space-time coordinates
 are doubled by
 introducing winding type coordinates, which combine with
 the space-time coordinates to form an $O(d,d)$ vector $X^M = (\tilde{x}_{\mu} ,
 x^{\mu})$ that transforms as $X^M
\rightarrow X'^M = T^M_{\ N} X^N$ (see section \ref{DFTreview}).
This implies $\partial_M \rightarrow \del'_M =  (T^{-1})_M^{\ N}
\del_N$.
  Now, using $S^{-1} \Gamma^M S =
T^M_{\ P} \Gamma^P,$ which follows directly from $\rho(S) = T$,
one can show \bea \label{constantS} \slashed{\partial}\chi &=&
\Gamma^M \partial_M \chi
  \rightarrow
\slashed{\partial}'\chi' = \Gamma^M (T^{-1})^N_{\ M} \del_N (S
\chi) \nonumber \\
 & = & S S^{-1} \Gamma^M S (T^{-1})^N_{\ M} \del_N \chi = S
\ \slashed{\partial}\chi. \eea The transformation of
$\slashed{\del} \chi$ implies that \be \label{finalF} F = e^{-B}
\slashed{\partial}\chi \rightarrow F' = e^{-B'} \slashed{\del}'
\chi' = e^{-B'} S \slashed{\partial}\chi = e^{-B'} S e^{B} F. \ee


\section{NATD  as an  $O(d,d)$ transformation}\label{NATD}

Non-Abelian T-duality can be applied by using the standard tools
of the Buscher method. For a generic nonlinear sigma model with
isometry group $G$, one starts with gauging the symmetry group (or
a subgroup of it) and introduces Lagrange multiplier terms which
constrains the gauge field to be pure gauge. Integrating out the
Lagrange multipliers, one obtains the original model. Integrating
out the gauge field gives the NATD model, for which the Lagrange
multiplier terms play the role of coordinates on the dual
manifold.

The NATD of a generic Green-Schwarz string sigma model with
isometry group $G$ has  been recently obtained in
\cite{Borsato:2018idb}. Here we present their results (for bosonic
$G$) and show that the new backgrounds can also be obtained by
applying a coordinate dependent $O(d,d)$ transformation embedded
in $O(10,10)$. The best way to present the rules for
transformation is to introduce coordinates which makes the
isometry symmetry manifest. With respect to such coordinates one
can write
\begin{eqnarray} ds^2 &=& G_{\mu \nu} dx^{\mu} dx^{\nu} = G_{mn}
dx^{m} dx^{n} + 2 G_{mi} dx^{m} d\theta^{i} + G_{ij} d\theta^{i}
d\theta^{j} \label{metric1} \\
& = & G_{mn} dx^{m} dx^{n} + 2 G_{mI} dx^{m} \sigma^I + G_{IJ}
\sigma^I \sigma^J \label{metric2} \\
& = & G_{\alpha \beta} \sigma^{\alpha} \sigma^{\beta},
\label{metric3}
\end{eqnarray} where $\theta^{i}, i= 1, \cdots, d$ are coordinates for $G$ and $\sigma^a = \delta^a_{\ m} dx^m$ and $\sigma^I, I = 1, \cdots, dimG $
are the left invariant 1-forms $\sigma^I = l^I_{\ i} d\theta^{i}$
on  $G$   defined from the Maurer-Cartan form: $g^{-1} dg =
\sigma^I T_I$ with $T_I$ forming a basis for the Lie algebra
${\cal{G}}$ of $G$. Similarly, \bea \label{B1} B &=& \frac{1}{2}
B_{\mu
\nu} dx^{\mu} \wedge dx^{\nu} \\
\label{B2} &=& \frac{1}{2} B_{mn} dx^{m} \wedge dx^{n} + B_{mI}
dx^{m} \wedge \sigma^I + \frac{1}{2} B_{IJ} \sigma^I \wedge
\sigma^J. \eea Since the group $G$ acts on the background by
isometries, all the $\theta$ dependence of the fields are encoded
in $l^I_{\ i}$. After applying NATD with respect to $G$, one ends
up with a sigma model which corresponds to the following
background\footnote{Note that due to the convention in
\cite{Borsato:2018idb} there is a difference in the sign in front
of the B field term.} \begin{eqnarray}
G'_{mn} & = & G_{mn} - [(G+B) N (G+B)]_{(mn)} \\
G'_{m I} & = & \frac{1}{2} [(G+B) N]_{m I} -  \frac{1}{2} [N
(G+B)]_{I m} \\
G'_{IJ} & = & N_{(IJ)} \\
B'_{mn} & = & B_{mn} - [(G+B) N (G+B)]_{[mn]} \\
B'_{m I} & = & -\frac{1}{2} [(G+B) N]_{m I} -  \frac{1}{2} [N
(G+B)]_{I m} \\
B'_{IJ} & = & -N_{[IJ]},
\end{eqnarray} where \be \label{defnofN}
N^{IJ} = (G_{IJ} + B_{IJ} + \nu_K C_{IJ}^{\ \ K})^{-1}. \ee Here,
$C_{IJ}^{\ \ K}$ are the structure constants of the Lie algebra
${\cal{G}}$ with respect to the basis $T_I$, that is, $[T_I, T_J]
= C_{IJ}^{\ \ K} T_K.$ The metric and the B-field in the
transformed background are
\begin{eqnarray} \label{newmetric} ds^2 &=& G'_{mn} dx^{m} dx^{n}
+ 2 G'_{m I} dx^{m} d\nu^I + G'_{IJ} d\nu^I d\nu^J \\
\label{newBfield}B' &=& \frac{1}{2} B'_{mn} dx^{m} \wedge dx^{n} +
B'_{m I} dx^{m} \wedge d\nu^I + \frac{1}{2} B'_{IJ} d\nu^I \wedge
d\nu_J.
\end{eqnarray} Here, $\nu_I$ are  Lagrange multiplier terms in the Buscher method. They parameterize the dual space and hence
they have lower indices as in (\ref{defnofN}). Those indices are
raised by the Kronecker delta $\nu^I = \delta^{IJ} \nu_J$ in
(\ref{newmetric}, \ref{newBfield}) so that they have the standard
upper placement of indices as coordinates of the NATD fields. This
has also been discussed in \cite{Borsato:2018idb}, see their
footnote 7.

The transformation for the dilaton field presented in
\cite{Borsato:2018idb}  is \be \label{dilaton2} \phi' = \phi +
\frac{1}{2} \ln {\rm det} N. \ee

Now, let us write the above transformation rules in the
terminology of the previous section. We define the background
matrix $Q = G + B$ and $Q' = G' + B'$. Then the above rules become
\begin{eqnarray}
Q'_{mn} &=& Q_{mn} - [(Q_{m I} N^{IJ} Q_{J n}] \\
Q'_{m I} &=& [Q N]_{m I} \\
Q'_{I m} &=& -[N Q]_{I m} \\
Q'_{IJ} &=& N_{IJ} \end{eqnarray}

Comparing this with (\ref{genelQ}) and (\ref{genelE}) one
immediately sees that the new background has been obtained by the
action of the fractional linear transformation with the following
$O(d,d)$ matrix $T_{{\rm NATD}}$ embedded in
$O(10,10)$\footnote{We name both matrices (the $O(d,d)$ matrix
(\ref{NATDmatrix}) and the the $O(10,10)$ matrix in which it is
embedded) as $T_{{\rm NATD}}$.} in the way presented in the
section above: \be \label{NATDmatrix} T_{{\rm NATD}} =
\left(\begin{array}{cc} 0 & 1 \\
                        1 & \theta_{IJ} \end{array}\right), \ \ \ \
                        \theta_{IJ}
                        = \nu_K C_{IJ}^{\ \ K}. \ee
Let us also check that the transformation rule (\ref{dilaton2})
for the dilaton field can be obtained through the action of
$T_{{\rm NATD}}$ by comparing it with (\ref{dilaton1}). It is a
well known fact that the transformation (\ref{finalQ}) implies for
$G'$ the following \cite{Giveon:1994fu}: \be G' =
\frac{1}{(\hat{c} Q + \hat{d})^T} \ G \ \frac{1}{(\hat{c} Q +
\hat{d})}. \ee Then, \be \frac{{\rm det} G'}{{\rm det} G} =
\left({\rm det}(\hat{c} Q + \hat{d})^{-1}\right)^2. \ee  When $T$
is as in (\ref{NATDmatrix}) this gives \be \sqrt{\frac{{\rm det}
G'}{{\rm det} G}} = {\rm det} N, \ee and the two expressions
(\ref{dilaton1}) and (\ref{dilaton2}) indeed match.

It is important to note that the dimension $d$ of the isometry
group determines whether the the NATD matrix $T_{{\rm NATD}}$ acts
within Type IIA/Type IIB or it involves a reflection which implies
that a Type IIA solution is mapped to a Type IIB solution or vice
versa. The former situation arises when $d$ is even and the latter
occurs when $d$ is odd.

Since we have identified the $O(10,10)$ matrix that generates the
NS-NS sector of the NATD background, we can immediately determine
the transformed RR sector, as well. All we have to do is to find
the $Pin(10,10)$ matrix that acts on the spinor field that
packages the modified p-form gauge potentials in the democratic
formulation.  The $Pin(10,10)$ element $S_{{\rm NATD}}$ that
projects to the $O(10,10)$ element (\ref{NATDmatrix}) under the
double covering homomorphism $\rho : Pin(d,d) \rightarrow O(d,d)$
can be found easily: \be S_{{\rm NATD}}= C S_{\theta} = S_{\beta}
C, \ee where $C$ is the charge conjugation matrix. For more
details, see \cite{dftRR} and \cite{Aybike}. The factors
$S_{\theta}$ and $S_{\beta}$ in $S_{{\rm NATD}}$ are the
$Spin^+(10,10)$ elements that projects onto the $SO^+(10,10)$
matrix that generates the $B$-transformations and $\beta$-shifts
with $\theta_{IJ}
                        = \nu_K C_{IJ}^{\ \ K}$ and $\beta_{IJ}
                        = \nu_K C_{IJ}^{\ \ K}$, respectively. Then
the transformation of the p-form fluxes is \be \label{finalF2} F'
= e^{-B'} S_{{\rm NATD}} e^{B} F. \ee

An important remark is in order here. Recall from the discussion
in section (\ref{RRsector}) that the transformation (\ref{finalF})
is equivalent to the transformation (\ref{constantS}). Also recall
that the transformation (\ref{constantS})  equivalent to the
transformation (\ref{transfchi}), when $S$ is constant. However,
when $S$ is not constant as in here, the two transformations
(\ref{constantS}) and (\ref{transfchi}) are not equivalent.
Naively, one would have expected that the right transformation
rule for the RR fields under NATD would follow from the
transformation \be \chi \rightarrow  S_{{\rm NATD}} \chi, \ee
which would imply a different transformation rule for the field
strength $F$ that would also involve fluxes associated with
$S_{{\rm NATD}}$ (see section (\ref{fluxes})). However, the right
transformation rule is as in (\ref{finalF2}), as we will
demonstrate in the next section through the example of the $AdS_3
\times S^3 \times T^4$. Then in section (\ref{eom}), we will prove
that the transformed fields will constitute a solution of the GSE
when the transformation for the RR fields is as in
(\ref{finalF2}).


\subsection{An Example: $AdS_3 \times S^3 \times T^4 $}\label{example}

Let us consider the simple example $AdS_3 \times S^3 \times T^4$.
This geometry arises as the near horizon limit of the D1-D5
system. The geometry has to be supported by 3-form Ramond-Ramond
flux. We have
\begin{eqnarray} ds^2 & =
& ds^2(AdS_3) + ds^2(S^3) + ds^3(T^4) \\
F_3 & = & Vol(S^3) + Vol(AdS_3)
\end{eqnarray}
Note that we also need the Hodge dual of the 3-form flux which is
the following 7-form flux: \be F_7 = -(\star F_3) = (Vol(S^3) +
Vol(AdS_3)) \wedge Vol(T^4). \ee

Due to the presences of the 3-sphere in the geometry, one has a
global $SO(4) \simeq SU(2) \times SU(2)$ isometry symmetry. It is
possible to use one of these $SU(2)$ groups to apply NATD. Writing
the $S^3$ part of the metric as \be ds^2(S^3) = (\sigma^1)^2 +
(\sigma^2)^2 + (\sigma^2)^2, \ee where $\sigma^I, \ I=1,2,3$ are
the 3 left invariant 1-forms for $SU(2)$, we have \be Q_{m I} =
Q_{I m} = 0 \ee and \be Q_{IJ} = E_{IJ} = G_{IJ} = \delta_{IJ}.
\ee Now we apply the NATD matrix (\ref{NATDmatrix}) on this
background. Here, the structure constants that determine the NATD
matrix are $C_{IJ}^{\ \ K} =\epsilon_{IJ}^{\ \ K}$. This gives
\begin{eqnarray} ds^2(S^3_{{\rm def}}) & = &
\frac{1}{1 + r^2} \sum_{I,J=1}^3 \left((1+ \nu^I)^2 (d\nu^I)^{2} +
2
\nu^I \nu^J d\nu^I d\nu^J \right)  \\
B & = & \frac{1}{1 + r^2} \left(-\nu^3 d\nu^1 \wedge d\nu^2 +
\nu^2 d\nu^1 \wedge d\nu^3 - \nu^1 d\nu^2 \wedge d\nu^3 \right),
\label{AdS3Bfield}
\end{eqnarray}
where $r^2 = (\nu^1)^2 + (\nu^2)^2 + (\nu^3)^2$. Writing this in
spherical coordinates
\begin{equation}
\nu^1 = r \sin{\theta} \cos{\phi} \ \ \ \ \nu^2  = r \sin{\theta}
\sin{\phi} \ \ \ \ \nu^3  = r \cos{\theta}
\end{equation}
we have
\begin{eqnarray} ds^2 & =
& ds^2(AdS_3) + ds^2(\tilde{S}^3) + ds^3(T^4)\nonumber  \\
\label{AdS3metric} ds^2(\tilde{S}^3) & = &dr^2 + \frac{r^2}{1+r^2} d\Omega^2 = dr^2 + \frac{r^2}{1+r^2}\left(d\theta^2 + \sin^2{\theta} d\phi^2\right)  \\
\nonumber B' & = & - \frac{r^3}{1+r^2} Vol(S^2) = - \frac{r^3}{1+r^2} \sin{\theta} d\theta \wedge d\phi \\
\Phi'& = & -\frac{1}{2} \ln{(1+r^2)}
\end{eqnarray}

Now, let us look at the transformation of the RR sector. Similar
to the Abelian case we form the differential form, which encodes
the RR fluxes \be F = \sum_p G^{(p)} = \sum_p \left(F^{(p)} +
F_I^{(p-1)} \sigma^I + \frac{1}{2} F_{IJ}^{(p-2)} \sigma^I \wedge
\sigma^J +  F^{(p-3)} \sigma^1 \wedge \sigma^2 \wedge \sigma^3
\right) \ee where we have decomposed a $p-$form RR flux $G^{(p)}$
according to how many legs it does have along the directions of
the isometry group $SU(2)$. The fluxes $F^{(p-a)}, \ a=0,1,2,3$
have no dependence on the coordinates $r, \theta, \phi$. We map
this differential form to a Clifford algebra element in the usual
way. The difference we have here is that it is $\sigma^I$ and not
$dx^i$ that we identify with the Clifford algebra element
$\psi^I$, for $I= 1,2,3$. On the other hand, for $a =d+1, \cdots,
10$, $dx^{a}$ is replaced with $\psi^{a}$, as usual. Here,
$\psi^{\alpha}, \ \alpha = (I, a)$ are the Clifford algebra
elements $\psi^{\alpha} = 1/\sqrt{2} \Gamma^{\alpha}$, where
$\Gamma^{\alpha}$ are the Gamma matrices. For more details, see
\cite{Aybike}. For index conventions, see Appendix
(\ref{conventions}).
 For the example we consider in this
section we only have 3- and 7-form fluxes, so the spinor field
takes the following form: \bea F &=& \psi^1 . \psi^2
. \psi^3 + F^{(3)} \psi^{\hat{1}}.\psi^{\hat{2}}.\psi^{\hat{3}} \\
&+& F^{(3)} F^{(4)} \psi^{\hat{1}}.\psi^{\hat{2}}.\psi^{\hat{3}}
\psi^{\hat{4}}.\psi^{\hat{5}}.\psi^{\hat{6}} \psi^{\hat{7}} +
F^{(4)} \psi^1 . \psi^2 . \psi^3
\psi^{\hat{4}}.\psi^{\hat{5}}.\psi^{\hat{6}} \psi^{\hat{7}}. \eea
Here, $Vol(AdS_3) = F^{(3)} dx^{\hat{1}} \wedge dx^{\hat{2}}
\wedge dx^{\hat{3}}$ and $Vol(T^4)= F^{(4)} dx^{\hat{4}} \wedge
dx^{\hat{5}} \wedge dx^{\hat{6}} \wedge dx^{\hat{7}}.$ Here, the
hatted numbers count the non-isometric directions. Note that
$F^{(3)}$ and $F^{(4)}$ are functions, not forms. Now we calculate
$F'$ from (\ref{finalF2}). First note that $ e^{B} F =
 F$, since the B-field is zero on the original background. Let us
 first calculate $S_{\theta} F = (1 + \nu_K \epsilon_{IJ}^{\ \ K}
 \psi^I
 . \psi^J) . F$. As one can easily calculate, this gives \bea
 S_{\theta} . F & = & F + \nu_K \epsilon_{IJ}^{\ \ K}
 \psi^I
 . \psi^J . F \nonumber \\ \label{step1}
 & = & F + F^{(3)} \nu_K \epsilon_{IJ}^{\ \ K}
 \psi^I
 . \psi^J .
\psi^{\hat{1}}.\psi^{\hat{2}}.\psi^{\hat{3}} (1 + F^{(4)}
\psi^{\hat{4}}.\psi^{\hat{5}}.\psi^{\hat{6}} \psi^{\hat{7}}). \eea
Now we apply the charge conjugation operator \cite{dftRR} \be C
\equiv (\psi^1 - \psi_1) . (\psi^2 - \psi_2) . (\psi^3 - \psi_3)
\ee on (\ref{step1}): \bea  C S_{\theta} . F &=& C . F - F^{(3)}
\nu_K \psi^K \psi^{\hat{1}}.\psi^{\hat{2}}.\psi^{\hat{3}} (1 +
F^{(4)} \psi^{\hat{4}}.\psi^{\hat{5}}.\psi^{\hat{6}}
\psi^{\hat{7}}) \nonumber \\  &=& 1 + F^{(4)}
\psi^{\hat{4}}.\psi^{\hat{5}}.\psi^{\hat{6}} \psi^{\hat{7}} +
F^{(3)} \psi^1 . \psi^2 . \psi^3 .
\psi^{\hat{1}}.\psi^{\hat{2}}.\psi^{\hat{3}} (1 + F^{(4)}
\psi^{\hat{4}}.\psi^{\hat{5}}.\psi^{\hat{6}} \psi^{\hat{7}})
\nonumber \\ \label{step2} && - F^{(3)} \nu_K \psi^K
\psi^{\hat{1}}.\psi^{\hat{2}}.\psi^{\hat{3}} (1 + F^{(4)}
\psi^{\hat{4}}.\psi^{\hat{5}}.\psi^{\hat{6}} \psi^{\hat{7}}).\eea

Finally, we apply $e^{-B'} = 1 + \frac{1}{1+r^2} \nu_K
\epsilon_{IJ}^{\ \ K} \psi^I . \psi^J $ on (\ref{step2}), where we
read off $B'$ from (\ref{AdS3Bfield}): \bea F' = e^{-B'} C
S_{\theta} F & = & 1 + F^{(4)}
\psi^{\hat{4}}.\psi^{\hat{5}}.\psi^{\hat{6}} \psi^{\hat{7}}
\nonumber \\ & & + (1 - \frac{r^2}{1 + r^2}) F^{(3)}\psi^1 .
\psi^2. \psi^3 . \psi^{\hat{1}}.\psi^{\hat{2}}.\psi^{\hat{3}} (1 +
F^{(4)} \psi^{\hat{4}}.\psi^{\hat{5}}.\psi^{\hat{6}}
\psi^{\hat{7}}) \nonumber \\ & & - \nu_K \psi^K F^{(3)}
\psi^{\hat{1}}.\psi^{\hat{2}}.\psi^{\hat{3}} (1 + F^{(4)}
\psi^{\hat{4}}.\psi^{\hat{5}}.\psi^{\hat{6}} \psi^{\hat{7}})
\nonumber \\ & & + \frac{1}{1 + r^2} \nu_K \epsilon_{IJ}^{\ \ K}
\psi^I . \psi^J . (1 + F^{(4)}
\psi^{\hat{4}}.\psi^{\hat{5}}.\psi^{\hat{6}} \psi^{\hat{7}}).
\label{sonuc}
 \eea
From $F'$ we can read off the p-form fluxes of the dual background
after now identifying $\psi^I$ with $d\nu^I$. Since it is only the
$S^3$ directions that have been dualized, we still have $$F^{(4)}
\psi^{\hat{4}}.\psi^{\hat{5}}.\psi^{\hat{6}} \psi^{\hat{7}}
\leftrightarrow Vol(T^4) \ {\rm and} \  F^{(3)}
\psi^{\hat{1}}.\psi^{\hat{2}}.\psi^{\hat{3}} \leftrightarrow
Vol(AdS_3),$$  The fluxes in the NATD background are then found as  \bea F_0 & = & 1 \nonumber \\
                                      F_2 & = & \frac{1}{1+r^2} (\nu^1 d\nu^2 \wedge d\nu^3 + \nu^2 d\nu^3 \wedge d\nu^1 + \nu^3
d\nu^1 \wedge d\nu^2) =  \frac{r^3}{1+r^2} Vol(S^2) \nonumber  \\
                                      F_4 & = & Vol(T^4) - \sum_{I=1}^3 \nu^I d\nu^I Vol(AdS_3) =  Vol(T^4) - r dr Vol(AdS_3)  \nonumber   \\
                                      F_6 & = & \frac{1}{1+r^2} \left[d\nu^1 \wedge d\nu^2 \wedge d\nu^3 \wedge Vol(AdS_3)
                                      + (\nu^1 d\nu^2 \wedge d\nu^3 + \nu^2 d\nu^3 \wedge d\nu^1 + \nu^3
d\nu^1 \wedge d\nu^2) Vol(T^4)\right]  \nonumber \\
& = &\frac{r^2}{1+r^2} dr \wedge Vol(S^2) \wedge Vol(AdS_3) + \frac{r^3}{1+r^2} Vol(S^2) \wedge Vol(T^4) \nonumber \\
& = & Vol(\tilde{S}^3) \wedge Vol(AdS_3) + \frac{r^3}{1+r^2}
Vol(S^2) \wedge Vol(T^4) \nonumber \\
                                      F_8 & = & - \sum_{I=1}^3 \nu^I d\nu^I \wedge Vol(AdS_3) \wedge Vol(T^4) =  -r dr \wedge Vol(AdS_3) \wedge Vol(T^4) \nonumber  \\
                                      F_{10} & = & Vol(\tilde{S}^3)
                                      \wedge Vol(AdS_3) \wedge
                                      Vol(T^4) = \star 1. \nonumber \eea

 Here, \be Vol(\tilde{S}^3) = \frac{r^2 \sin{\theta}}{1+r^2} dr \wedge
d\theta \wedge d\phi = \frac{1}{1+r^2} Vol(S^3) = \frac{r^2}{1 +
r^2} dr Vol(S^2), \nonumber \ee and $\star$ is the Hodge
                                      dual with respect to the
                                      metric of the deformed
                                      background given in (\ref{AdS3metric}).
                                      These results match exactly with the
                                      results obtained in \cite{Sfetsos:2010uq} by
                                      conventional methods of
                                      NATD.


\section{NATD as a solution generating transformation in
Double Field Theory}\label{DFT}

The purpose of this section is to show that the NATD fields
obtained by applying the transformation (\ref{finalQ})  and
(\ref{finalF2}), where $T$ in (\ref{matrixT}) is the NATD matrix
 (\ref{NATDmatrix}) are solutions of (generalized) supergravity
equations. We find it  useful to discuss this in the framework of
Double Field Theory (DFT), where $O(d,d)$ arises as a manifest
symmetry of the action and hence of the field equations.
Therefore, we start with a brief review of DFT.

\subsection{A Brief Review of Double Field Theory}\label{DFTreview}

DFT is a field theory defined on a doubled space, which implements
the $O(d,d)$ T-duality symmetry of string theory as a  manifest
symmetry. In addition to the standard space-time coordinates, the
doubled space also includes dual coordinates, which are associated
with the winding excitations of closed string theory on
backgrounds with non-trivial cycles. The space-time and the dual
coordinates transform as a vector under the T-duality group
$O(d,d)$: \be X'^M =
h^M_{\ \ N} X^N, \ \ \ \ X^M =  \left(\ba{c} \tilde{x}_{\mu} \\
                                              x^{\mu} \ea \right) \ee
Here $\tilde{x}_{\mu}$ are the  dual coordinates and $h^M_{\ \ N}$
is a general $O(d,d)$ matrix. In what follows we will always
decompose the indices $M$ labelling the $O(d,d)$ representation as
$^M = (_{\mu}, \ ^{\mu})$, where $^{\mu}$ and $_{\mu}$ label
representations of the $GL(d)$ subgroup of $O(d,d)$. We will raise
and lower indices by the $O(d,d)$ invariant metric $\eta$, so that
$X_M = \eta_{MN} X^N$.

In DFT, the dynamical fields are the fields $\cH, \ciftS, d$ and
$\chi$. They are all allowed to depend on both the standard and
the winding type coordinates. The generalized metric $\cH$ is an
element of $SO^-(d,d)$ and it encodes the semi-Riemannian metric
and the B-field, see (\ref{genmet}). The generalized dilaton field
$d$ is defined from $e^{-2 d} = \sqrt{g} e^{-2 \phi}$ and it is
$O(d,d)$ invariant as was discussed in section \ref{NSsector}, see
(\ref{gendilaton2}).  The spinor field $\ciftS$  is the element in
$Spin^-(d,d)$ that projects onto $\cH$ under the double covering
homomorphism (\ref{coveringmap}) between $Spin(d,d)$ and
$SO(d,d)$, that is $\rho(\ciftS) = \cH$. The spinor field $\chi$
encodes the RR fields in the democratic formulation of Type II
supergravity. For more details see \cite{dftRR}, \cite{Aybike}.

The DFT action is as below: \be {\cal{S}} = \int \ dx d\tilde{x}
\left( {\cal{L}}_{{\rm NS-NS}} + {\cal{L}}_{{\rm RR}} \right), \ee
where \be \label{DFTaction1} {\cal{L}}_{{\rm NS-NS}} = e^{-2 d} \
{\cal{R}} (\cH, d) \ee and
\begin{equation}\label{DFTaction} {\cal{L}}_{{\rm RR}} =
\frac{1}{4} \langle  \ \slashed{\partial} \chi, \ C^{-1} \ciftS \
\slashed{\partial} \chi \rangle.
\end{equation} Here, $\langle ~ \rangle$ is the Mukai pairing,
which is a $Spin(d,d)$ invariant bilinear form on the space of
spinors \cite{Mukai}. This action has to be implemented by the
following self-duality condition \be \label{selfduality}
\slashed{\partial} \chi = - \cK \ \slashed{\partial}\chi, \ \ \
\cK \equiv C^{-1} \ciftS. \ee Moreover, one needs to impose the
following $O(d,d)$ covariant constraint, which is called the
strong constraint:
  \be\label{ODDconstr}
   \partial^{M}\partial_{M}A \ = \ \eta^{MN}\partial_{M}\partial_{N}A \ = \ 0\;, \qquad
   \partial^{M}A\,\partial_{M}B \ = \ 0\;, \qquad
   \eta^{MN} \ = \  \begin{pmatrix}
    0&1 \\1&0 \end{pmatrix}\,,
 \ee
where $A$ and $B$ represent any fields or parameters of the
theory. When the constraint is satisfied, the DFT action is gauge
invariant under generalized diffeomorphisms and the gauge algebra
closes under the C-bracket, which is an $O(d,d)$ covariant
extension of the Courant bracket. When the fields and the gauge
parameters have no dependence on the winding type coordinates,
that is, when $\tilde{\partial}^{\mu} = 0$, the strong constraint
(\ref{ODDconstr}) is satisfied trivially. In this case, the theory
is said to be in the supergravity frame because for this solution
of the constraint it can be shown that (\ref{DFTaction1}) reduces
to the standard NS-NS action for the massless fields of string
theory and (\ref{DFTaction}) reduces to the RR sector of the
democratic formulation of Type II supergravity theory.

The term $\cR(\cH, d)$ in (\ref{DFTaction1}) is the generalized
Ricci scalar and its explicit form is as follows: \bea
\label{ricci} \cR(\cH, d) & = & 4 \cH^{MN}
\partial_M
\partial_N d - \del_M \del_N \cH^{MN} - 4 \cH^{MN} \del_M d \del_N d + 4 \del_M \cH^{MN} \del_N d \\
& & + \frac{1}{8} \cH^{MN} \del_M \cH^{KL} \del_N \cH_{KL} -
\frac{1}{2} \cH^{MN} \del_M \cH^{KL} \del_K \cH_{NL} \nonumber
\eea

The DFT action presented in (\ref{DFTaction}) is invariant under
the following transformations: \be \label{transform}
 \ciftS(X) ~~  \longrightarrow \ciftS^{\prime}(X^{\prime}) \ = \ (S^{-1})^{\dagger}\, \ciftS(X)
 \,S^{-1}\;, ~~ \chi(X) \longrightarrow \chi(X') = S \chi(X)
 \ee
Here $S \in Spin(d,d)$  and $X^{\prime}= hX$,  where $h = \rho(S)
\in SO^+(d,d)$. As mentioned before, the generalized dilaton field
is $O(d,d)$ invariant. The transformation rules for the
generalized metric $\cH = \rho(\ciftS)$ is determined by those of
$\ciftS$ and is as given below: \be \label{transform2}
 \cH(X) ~~  \longrightarrow \cH^{\prime}(X^{\prime}) \ = \ (h^{-1})^{T}\, \cH(X)
 \,h^{-1}\;.
 \ee The generalized Ricci scalar (\ref{ricci}) is manifestly
 invariant under these transformations. A fact that is of crucial
 importance is that the transformation (\ref{transform2}) is
 equivalent to \cite{ZwiebachL} \be Q \rightarrow Q' = (A Q + B)(C Q + D)^{-1}, \ee
 where \be h = \left(\begin{array}{cc} A & B \\
                        C & D \end{array}\right). \nonumber \ee

\subsection{Embedding NATD in Double Field Theory}

 We showed in the previous section that the NATD of a given Type II background with isometry $G$ can be
obtained through the action of the $O(d,d)$ matrix
(\ref{NATDmatrix}).  As we have mentioned before, the field
equations of DFT reduce to the field equations of Type II
supergravity for the trivial solution of the constraint, that is
when the fields are in the supergravity frame. As a result, the
type II supergravity solution on which the NATD acts can  also be
regarded as a a solution for DFT. Now, assume that the  isometry
group $G$ is unimodular\footnote{We will relax the condition of
unimodularity later on.} and that it acts freely on the
background. The latter condition means that one can pick up
coordinates in which the metric and the B-field can be written as
in (\ref{metric2}) and (\ref{B2}). We label these coordinates as
$\{x^1, \cdots, x^{10-d}, \theta^1, \cdots, \theta^d \}$; then the
dual coordinates will be labelled as $\{\tilde{x}^1, \cdots,
\tilde{x}^{10-d}, \tilde{\theta}^1, \cdots, \tilde{\theta}^d \}$.
Obviously, the DFT fields $\cH, \ciftS, d$, $\chi$ that correspond
to this background do not depend on the dual coordinates, that is,
they are in the supergravity frame.

Since the group $G$ acts on the background by isometries, all the
$\theta$ dependence of the fields in (\ref{metric2}) and
(\ref{B2}) are encoded in $l^I_{\ i}$. We define the matrices
$G(x, \theta), G(x), B(x, \theta)$ and $B(x)$ from \be
\label{denklem1} ds^2 = d\textbf{x}^T G(x, \theta) d\textbf{x} =
\sigma^T G(x) \sigma, \ \ \ \ B = d\textbf{x}^T B(x, \theta)\wedge
d\textbf{x} = \sigma^T B(x) \wedge \sigma, \ee where $\wedge$
denote the obvious wedge product of matrices and $d\textbf{x}$ and
$\sigma$ denote the 10-vectors with components $(dx^1, \cdots,
dx^{10})$ and $(\sigma^1, \cdots, \sigma^d, dx^{d+1}, \cdots,
dx^{10})$, respectively. Then the background matrix  $Q = G + B$
in (\ref{background}) has the following form: \be Q(x, \theta) =
l^T(\theta) Q(x) l(\theta), \ee where $l$ is the $GL(10)$ matrix
obtained by embedding the  $GL(d)$ matrix $l_d$ with components
$(l_d)^I_{\ i} = l^I_{\ i}$. The embedding is as described in
(\ref{embedding}), so $(l_d)^I_{\ m} = l^a_{\ i} = 0$ and
$(l_d)^a_{\ m} = \delta^a_{\ m}$. This is equivalent to the
following $O(10,10)$ action : \be \label{separatedE}
 Q(x, \theta) =L(\theta). Q(x),\ee where $L$ is the  $O(10,10)$
 matrix \be \label{geomtwist} L = \left(\begin{array}{cc}
                                  l^T & 0 \\
                                  0 & l^{-1}
                                  \end{array}\right). \ee
As stated at the end of section \ref{DFTreview}, the equation
(\ref{separatedE}) is equivalent to \cite{ZwiebachL}:\be
\label{separatedH}  \cH(x, \theta)= L(\theta) \cH(x) L^T(\theta).
\ee Hence, the dependence of the generalized metric $\cH$ on the
coordinates $(x, \theta)$ is separated. Since the twist matrix $L$
operates between curved and flat indices, the index structure of
it is as follows: \be \label{separatedH2} \cH^{MN}(x^1, \cdots,
x^{10-d}, \theta^1, \cdots, \theta^d) = L^M_{\ \ A}(\theta^1,
\cdots, \theta^d) \cH^{AB}(x^1, \cdots, x^{10-d}) L^N_{\ \
B}(\theta^1, \cdots, \theta^d), \ee where we have identified \be
\label{genmet} \cH \longleftrightarrow \cH^{MN}
\longleftrightarrow
\left(\begin{array}{cc} G - B G^{-1} B &  G^{-1} \\
                        - G^{-1} B & G^{-1} \end{array}\right).
                        \ee From (\ref{geomtwist}) we read off
                        $L_m^{\ a} = \delta_m^{\ a}, \ L^m_{\ a} =
                        \delta^m_{\ a}, \ L_m^{\ I} =  L^m_{\ I} =
                        L_i^{\ a} = L^i_{\ a} = 0$ and $L_i^{\ I}=
                        (L_d)_i^{\ I} = l_i^{\ I}, \ L^i_{\ I}
                        =(L_d)^i_{\ I} = l^i_{\ I}$, where $l^i_{\
                        I} l^I_{\ j} =  \delta^i_{\ j}$.

Similarly, the dependence of the field $\ciftS$  on the
coordinates $(x, \theta)$ is also separated. \be
\label{separatedS} \ciftS(x^1, \cdots, x^{10-d}, \theta^1,
\cdots,\theta^d ) = (S_L^{-1})^{\dagger}(\theta^1,
\cdots,\theta^d) \ciftS(x^1, \cdots, x^{10-d})
(S_L)^{-1}(\theta^1, \cdots,\theta^d)\ee Here,  $S_L$ is the
$Pin(10,10)$ matrix that projects onto $L$ under the double
covering homomorphism: $\rho(S_L) = L$. For $\cK \equiv C^{-1}
\ciftS $, this implies \be \label{anzatsK} \cK(x, \theta) =
S_L(\theta) \cK(x) S_L^{-1}(\theta). \ee

We also assume that the p-form field strengths (not the gauge
potentials) respect this isometry, that is, we assume that any
p-form flux in the background can be written as \be
\label{spinorF} G^{(p)} = \sum_p \left(F^{(p)}(x) + F_I^{(p-1)}(x)
\sigma^I + \frac{1}{2} F_{IJ}^{(p-2)}(x) \sigma^I \wedge \sigma^J
+ \cdots + F^{(p-d)} \sigma^1 \wedge \cdots \wedge \sigma^d
\right). \ee Here, we have decomposed a $p-$form RR flux $G^{(p)}$
according to how many legs it does have along the directions of
the isometry group $G$. Since $G$ acts by isometries, the fluxes
$F^{(p-a)}(x), \ a=0,1,\cdots, d$ will have no dependence on the
isometry coordinates $\theta^i$. Let  $F$ be the differential form
that packages these p-forms as in the democratic formulation: \be
F = \sum_p G^{(p)}. \ee This can be regarded as a spinor field as
discussed in section \ref{NATD}, for more details, see
\cite{dftRR} and \cite{Aybike}. Owing to the form (\ref{spinorF}),
we have  \be \label{separatedF1} F(x, \theta) = S_L(\theta) F(x)
\ee where $F(x, \theta)$ is the spinor field that encodes the
components of the field strengths written with respect to the
coordinate basis $(dx^1,\cdots, dx^{(10-d)}, d\theta^1,\cdots,
d\theta^d)$. As we will show in a moment, when the twist matrix is
of the form (\ref{geomtwist}), (\ref{separatedF1}) is equivalent
to the following \be \label{separatedF} F(x, \theta) = e^{-B(x,
\theta)} S_L(\theta) e^{B(x)} F(x). \ee As we have mentioned
before, the
 relation (\ref{separatedF}) above is equivalent to \be
\label{separatedchi} \slashed{\del}\chi(x^1, \cdots, x^{n-d},
\theta^1, \cdots,\theta^d )= S_L (\theta^1, \cdots,\theta^d)
\slashed{\del}\chi(x^1, \cdots, x^{n-d}). \ee In order to show the
equivalence of the equations (\ref{separatedF1}) and
(\ref{separatedF}), first note that \be \label{denklem2}
S_L^{-1}(\theta) e^{B(x, \theta)} S_L(\theta) = e^{B(x)}. \ee This
follows from (\ref{denklem1}), which implies that $B(x, \theta) =
l^T B(x) l$. Writing \be h_B = \left(\begin{array}{cc} 1 & B \\
                                    0 & 1 \end{array}\right), \ee this means \be L^{-1} h_{B(x,\theta)} L = h_{B(x)}. \ee
Then, we have \be \rho(S_{L}^{-1}) \rho(e^{B(x,\theta)}) \rho(S_L)
= \rho(e^{B(x)}), \ee where $\rho$ is the double covering
homomorphism $\rho : Spin(d,d) \rightarrow O(d,d)$. Note that we
have used $\rho(e^B) = h_B$ and $\rho(S_L) = L$. Now, $\rho$ is a
homomorphism so the left hand side can be rewritten as
$\rho(S_L^{-1} e^{B(x,\theta)} S_L)$. This then gives
(\ref{denklem2}), as desired. Using this we immediately get \be
e^{-B(x, \theta)} S_L(\theta) e^{B(x)} F(x)= e^{-B(x, \theta)}
S_L(\theta) \underbrace{S_L^{-1}(\theta) e^{B(x, \theta)}
S_L(\theta)} F(x) = S_L(\theta) F(x), \ee where the the indicated
terms is written by using (\ref{denklem2}).

Using the terminology from duality twisted (Scherk-Shwarz)
reduction that we will discuss in subsection (\ref{sgdft}), we
call the fields $\cH(x), \ciftS(x), d(x)$ and $F(x)$
\emph{untwisted fields}.

Now, we apply the NATD transformation (\ref{finalQ})  and
(\ref{finalF2})  on these untwisted fields, where $T$ in
(\ref{matrixT}) is as in (\ref{NATDmatrix}). This will give us the
dual fields $\cH', d', \ciftS'$ and $F'$, which will depend on the
coordinates $\{x^1, \cdots, x^{10-d}, \nu^1, \cdots, \nu^d \}$,
which we collectively call $\{x, \nu\}$.
\begin{eqnarray}
\label{NATDH} \cH'^{MN}(x, \nu) &=& (T_{{\rm NATD}})^M_{\ \
A}(\nu)
\cH^{AB}(x) (T_{{\rm NATD}})^{N}_{\ \ B}(\nu)\\
\label{NATDS} \cK(x,  \nu ) &=& S_{{\rm NATD}}(\nu)
\cK(x) (S_{{\rm NATD}})^{-1}(\nu)\\
\label{NATDF} F'(x, \nu) &=& e^{-\sigma(\nu, \tilde{\nu})}
e^{-B'(x, \nu)} S_{{\rm NATD}}(\nu)
e^{B(x)} F(x) \\
\label{NATDd} d'(x, \nu) &=& d(x) + \sigma(\nu, \tilde{\nu}).
\end{eqnarray} Here, $\rho(S_{{\rm NATD}}) = T_{{\rm NATD}}$ and $B'(x, \nu) $ is read off from the antisymmetric part of $\cH'(x, \nu)$ in
(\ref{NATDH}). The field $\sigma(\nu, \tilde{\nu})$ in
(\ref{NATDd}) and (\ref{NATDF}) is non-vanishing only when the
isometry group is non-unimodular. We leave the discussion of this
term to section \ref{gse}.

Our strategy will be to show that these new fields $\cH'(x, \nu),
d'(x, \nu), \ciftS'(x, \nu)$ and $F'(x, \nu)$ form a solution for
the field equations of DFT.  Identifying  the coordinates $\{x,
\nu\}$ with the standard space-time coordinates, this means that
the corresponding supergravity fields in the NATD background form
a solution for the field equations of Type II supergravity, as the
field equations of DFT and Type II supergravity are equivalent in
the supergravity frame.

The key point in our argument will be to show that the two twist
matrices $L(\theta)$ and $T_{{\rm NATD}}(\nu)$ generate the same
fluxes defined in the framework of Gauged Double Field
Theory(GDFT). In the next section, we give a brief review of GDFT,
and introduce the fluxes that arise in this context. Finally, we
compute the fluxes associated with $L$ and $T_{{\rm NATD}}$ and
show that they are indeed the same.

\subsection{Gauged Double Field Theory and Fluxes Associated with the NATD matrix}

\subsubsection{Gauged Double Field Theory}\label{sgdft}

GDFT is obtained from duality twisted (Scherk-Schwarz) reduction
of DFT \cite{Geissbuhler}-\cite{Aybike}. The $O(d,d)$ invariance
of the DFT action under the transformations (\ref{transform}) and
(\ref{transform2}) makes it possible to introduce the following
Scherk-Schwarz type reduction anzats for the DFT fields: \be
\label{anzats1} \cH^{MN}(x, Y) = (U^{-1})^M_{\ \ A}(Y) \cH^{AB}(x)
(U^{-1})^N_{\ \ B}(Y), \ \ \cK(x, \theta ) = S(Y) \cK(x) S^{-1}(Y)
\ee
\begin{equation} \label{anzats2}  F(x, Y) = e^{-\sigma(Y)} e^{-B(x,Y)} S(Y)
e^{B(x)}F(x),
\end{equation} \be \label{anzats3} d(x, Y) = d(x) + \sigma(Y), \ee where $\rho(S) = U^{-1} \in O(d,d)$ and $F(x,Y) = e^{-B(x,Y)} \slashed{\del}\chi(x,Y)$. The matrices $U$ and
$S$ are usually called  twist matrices. When these anzatse are
plugged into the action and the gauge transformation rules of DFT,
all the $Y$ dependence is integrated out and one ends up with
GDFT, which is a consistent field theory for the untwisted fields
$\cH(x), \cK(x), d(x), F(x)$, provided that the matrices $U(Y),
S(Y)$ satisfy a set of constraints. We will list these constraints
in section \ref{fluxes}. The GDFT action is a deformation of the
DFT action determined by the so called fluxes, $f_{ABC}$. In the
NS-NS sector the Ricci scalar in (\ref{DFTaction1}) is deformed to
\be \label{deformedRicci} \cR \rightarrow \cR_{{\rm def}} = \cR +
\cR_f, \ee with \bea \label{deformedRicci2} \cR_f &=& -\frac{1}{2}
f^A_{\ BC} \cH^{BD} \cH^{CE} \del_D \cH_{AE} -\frac{1}{12} f^A_{\
BC} f^D_{\ EF} \cH_{AD} \cH^{BE} \cH^{CF}
\nonumber \\
& &  -\frac{1}{4} f^A_{\ BC} f^B_{\ AD} \cH^{CD} - 2 \eta_A \del_B
\cH^{AB} + 4 \eta_A \cH^{AB} \del_B d - \eta_A \eta_B \cH^{AB},
\eea The anzats in (\ref{anzats2}) does not yield  any deformation
in the GDFT action of the RR sector, as it is $F$ and not $\chi$,
which is twisted. As a result one ends up with the following
action\be \label{GDFT} S_{GDFT} = v \int dx \ d\tilde{x}
\left(e^{-2d} (\cR + \cR_f) + \frac{1}{4} \langle
\slashed{\del}\chi,
 C^{-1} \ciftS \  \slashed{\del}\chi \rangle\right) \ee where  $v$ is defined as \be \label{v}
v = \int d^d Y e^{-2\rho(Y)}. \ee  Explicit form of the fluxes
 that determine $\cR_f$  will be presented
 in the next subsection.
 The second term in (\ref{GDFT}) is the usual action for the RR sector
 of DFT of Type II strings and does not depend on the fluxes, as
  the duality twisted anzats has been imposed on the spinor field $F$ (which encode the RR fluxes), and not on the
 spinor field $\chi$ (which encodes the modified gauge potentials).
 Recall that the relation between the two is as in (\ref{ek1}). If it were the field $\chi$ which had been twisted, then the DFT action of the RR
sector would also be deformed in a way determined by the fluxes.
It was shown in \cite{Aybike} that the Lagrangian in this case is
of the same form as (\ref{DFTaction}), except that
$\slashed{\del}$ should be replaced with $\slashed{\nabla}$.
Although we will not need this deformed action in this paper, we
will need and present the explicit form of $\slashed{\nabla}$ in
section (\ref{dfteom}), where we will discuss the field equations
arising from (\ref{GDFT}).

\subsubsection{Fluxes, Dual Fluxes and the $O(d,d)$ invariance of
GDFT}\label{fluxes}

 The fluxes that determine the deformation in the NS-NS sector
are defined as below \cite{Grana} \be \label{structure} f_{ABC} =
3 \Omega_{[ABC]}, \ \ \ \eta_A =
\partial_M  (U^{-1})^M_{\ A} - 2(U^{-1})^M_{\ A} \partial_M \sigma \ee where $\sigma$ is as in (\ref{anzats3}) and \be \label{omega}
\Omega_{ABC} = -(U^{-1})^M_{\ A} \del_M (U^{-1})^N_{\ B} U^D_{\
N}\eta_{CD}.  \ee Note that $\Omega_{ABC}$ are antisymmetric in
the last two indices: $\Omega_{ABC} = - \Omega_{ACB}$. We also
make the following definition \be \label{fa} f_A = -\partial_M
(U^{-1})^M_{\ A}  = \Omega^C_{\ AC} \ee The constraints that
should be obeyed by the twist matrices are as follows: \be
\label{cond2}
 \partial^P (U^{-1})^M_{\ A} \partial_P g(X) = 0, \ee
\be \label{cond1}
 (U^{-1})^M_{\ A} \partial_M g(X) = \partial_A g(X). \ee
 where $g$ is
 any of the DFT fields ($\cH, \ciftS, \chi$).

 The DFT action of the NS-NS sector is manifestly $O(d,d)$
 invariant. So, if $h$ is a \emph{constant} $O(d,d)$ matrix, then
 \be \label{covofR} \cR[\cH, d, \del] = \cR[h^t \cH h, d, \hat{\del}]. \ee We have inserted  $\del$ and $\hat{\del}$ in the arguments of
$\cR$  to emphasize that  the derivatives $\del_M$ on the left
hand side should be replaced by $\hat{\del}_N \equiv h^M_{\ N}
\del_M$ on the right hand side. On the other hand, the DFT action
of the RR sector is  $Spin^+(d,d)$ invariant \cite{dftRR}.
Therefore, the field equations that arise from
 varying the DFT action of the NS-NS sector
 with respect to the generalized metric and the generalized dilaton are $O(d,d)$ covariant, whereas the
 field equations obtained by varying the DFT action of the RR sector with respect to the spinor
 field $\chi$ or the spinor field $\ciftS$  are covariant under the
 subgroup $Spin^+(d,d)$ of $Spin(d,d)$. This point will be
 important in
 section (\ref{dfteom}), see  equations (\ref{oddcov3}, \ref{oddcov4}).
 $Pin(d,d)$ elements that do not lie in this subgroup act as
 dualities rather than invariances, as we will discuss  in more
 detail in the next section. The $O(d,d)$ invariance of the generalized scalar curvature $\cR$ also extends to
 $\cR_f$, provided that  we treat the fluxes $f_{ABC}$ as spurious generalized
tensors, which also transform under $O(d,d)$. So, if we define \be
\label{spuriousf} \hat{f}^{ABC} = h^A_{\ D} h^{B}_{\ E} h^C_{\ F}
f^{DEF}, \ \ h \in O(d,d) \ee then it is easily shown that \be
\label{oddcov1} \cR_{\hat{f}}[h^t \cH h, d, \hat{\del}] =
\cR_f[\cH, d, \del]. \ee
 If the twist
matric $h^M_{\ \ N}$ satisfy the consistency condition
(\ref{cond1}) so that $\hat{\del}_M = \del_M$ acting on the fields
$\cH(x)$ and $d(x)$, then we simply have \bea \cR[h^t \cH h, d]
&=& \cR[\cH, d] \\ \cR_{\hat{f}}[h^t \cH h, d] &=& \cR_f[\cH, d].
\eea

At this point we find it useful to introduce \emph{dual fluxes}
and \emph{dual DFT fields}. Let us pick up $h = J$ in
(\ref{spuriousf}), where $J$ is the matrix obtained by embedding
the $d \times d$ matrix $J_d$ below in $O(10,10)$ as in
(\ref{embedding}): \be \label{jd} J_d =
\left(\begin{array}{cc} 0 & 1_d \\
1_d & 0 \end{array}\right).\ee In this particular case, we call
the resulting flux the \emph{dual flux}, and we denote it by
$\bar{f}^{ABC}$ for this particular case. That is, \be
\label{dualflux} \bar{f}^{ABC} = J^A_{\ D} J^{B}_{\ E} J^C_{\ F}
f^{DEF}.\ee   Note that, due to complete antisymmetry of $f_{ABC}$
in its indices, the only independent blocks of $f_{ABC}$ out of
the 8 possible combinations are $f^I_{\ JK}, f_{IJK}, f^{IJ}_{\ \
K}$ and $ f^{IJK}, \ I=1, \cdots, d$. It is customary to call
these the geometric flux, the H-flux, the Q-flux and the R-flux,
respectively \cite{Geissbuhler2}. Obviously, the geometric flux,
H-flux, Q-flux and R- flux components of $f$ is replaced by
Q-flux, R-flux, geometric flux and H-flux components in $\bar{f}$,
respectively. This is the reason why we  call the flux
$\bar{f}_{ABC}$ the dual flux of $f_{ABC}$. Now, taking $h = J$ in
(\ref{covofR}) and (\ref{oddcov1}), we get \be \label{oddcov2}
\cR[\bar{\cH}, d, \bar{\del}] = \cR[\cH, d, \del],\ \ \ {\rm and}
\ \ \ \cR_{\bar{f}}[\bar{\cH}, d, \bar{\del}] = \cR_f[\cH, d,
\del] \ee where we have defined \be \label{dual1} \bar{\cH} = J^t
\cH J, \ \ \bar{\del}_N = J^M_{\ \ N} \del_M . \ee The DFT field
$\bar{\cH}$ we defined above has been called the dual generalized
metric, in \cite{dftRR}.\footnote{Note that this is just a field
redefinition so there is no transformation on the coordinates. For
more details, see \cite{dftRR}.} Also note that we have $\del_i =
\tilde{\bar{\del}}^i, \ \tilde{\del}^i = \bar{\del}_i$, that is
the standard and dual derivatives have been swapped in $\del$ and
$\bar{\del}$. For future reference, we also define the dual spinor
fields $\bar{F}$ and $\bar{\chi}$ as in \cite{dftRR}: \be
\label{dual2} \bar{F} = e^{-\bar{B}} C e^{B} F, \ \ \ \bar{\chi} =
C \chi. \ee It is easily checked that $\bar{F} =e^{-\bar{B}}
\slashed{\del}\bar{\chi}$. Here, $\bar{B}$ is the B-field
associated with the dual generalized metric $\bar{\cH}$.

At this point, it is natural to ask the relation of the twist
matrix $\bar{U}$ associated with the fluxes $\bar{f}$ to the twist
matrix $U$ associated with the fluxes $f$. One can easily show
that the relation is $\bar{U} = J U$. Then, $\bar{S} =\pm  S C$,
where $\rho(\bar{S}) = \bar{U}^{-1}$ and $\rho(S) = U^{-1}$.
Indeed, it can easily be shown that \bea \label{omegabar}
 \bar{\Omega}_{ABC} &=& -(\bar{U}^{-1})^M_{\ A} \del_M
(\bar{U}^{-1})^N_{\ B} \bar{U}^D_{\ N}\eta_{CD} \\ \nonumber &=&
-J^D_{\ A} J^{E}_{\ B} J^F_{\ C}(U^{-1})^M_{\ D} \del_M
(U^{-1})^N_{\ E} U^G_{\ N}\eta_{FG} \\ &=& J^D_{\ A} J^{E}_{\ B}
J^F_{\ C} \Omega_{DEF}. \eea

For future reference, we also consider the fluxes associated with
the twist matrices $\breve{U} = U J$ and $\breve{S} = \pm C S $,
with $\rho(\breve{S}) = \breve{U}^{-1}$. One can easily see that
the fluxes associated with $\breve{U}$ are exactly the same as the
fluxes associated with $U$, except for the fact that all
standard/dual derivatives in the computation of $f$ should be
replaced with dual/standard derivatives in the computation of
$\breve{f}$. More precisely, we have \bea \label{omegatilde}
 \breve{\Omega}_{ABC} &=& -(\breve{U}^{-1})^M_{\ A} \del_M
(\breve{U}^{-1})^N_{\ B} \breve{U}^D_{\ N}\eta_{CD} \\ \nonumber
&=& -(U^{-1})^M_{\ A} \bar{\del}_M (U^{-1})^N_{\ B} U^D_{\
N}\eta_{CD}
\\ \nonumber&=& -(U^{-1})^M_{\ A} \del^M (U^{-1})^N_{\ B} U^D_{\
N}\eta_{CD}. \eea

\subsubsection{Fluxes associated with the NATD matrix}

 Let us now compute the fluxes associated with
 the matrices $L(\theta)$ in (\ref{geomtwist}) and $T_{{\rm NATD}}(\nu)$ in (\ref{NATDmatrix}). Note that the condition (\ref{cond1}) is trivially satisfied
 both by $L(\theta)$ and $T_{{\rm NATD}}(\nu)$,
 as they  are constructed
 by embedding $O(d,d)$ matrices in $O(10,10)$ as in
 (\ref{embedding}) and the $d$ coordinates on which $O(d,d)$ acts are not included in the $x$
 coordinates of the fields $\cH(x), d(x), \chi(x)$ and $\ciftS(x)$.  In the computation,  the coordinates on which the twist matrices
 depend are regarded as the standard coordinates and not the winding type ones. To be
 more precise, the $\theta$ coordinates of the geometric twist matrix $L(\theta)$
are the standard space coordinates for the fields in
 the supergravity background \emph{before} the dualisation. Then, after
 applying the NATD matrix $T_{{\rm NATD}}(\nu)$ on the untwisted fields, we end up with a set of fields, which
 now
 depend on the coordinates $(x, \nu)$. For the fields \emph{after}
 dualisation, it is now  these coordinates $(x, \nu)$ that are
 identified with the space-time coordinates. We start by expanding
 the formula given in (\ref{structure}):
 \bea \label{omega}
\Omega_{ABC} &=& -(U^{-1})^{\mu}_{\ A} \del_{\mu} (U^{-1})^N_{\ B}
U^D_{\ N}\eta_{CD} - (U^{-1})_{\mu A} \tilde{\del}^{\mu}
(U^{-1})^N_{\ B} U^D_{\
N}\eta_{CD} \nonumber \\
&=& -(U^{-1})^{i}_{\ A} \del_{i} (U^{-1})^N_{\ B} U^D_{\
N}\eta_{CD} \nonumber \\
&=& -(U^{-1})^i_{\ A} \del_i (U^{-1})^{j}_{\ B} U^D_{\ j}
\eta_{CD} -(U^{-1})^{i}_{\ A} \del_i (U^{-1})_{j B} U^{D j}
\eta_{CD}. \nonumber \eea In passing from the first line to the
second line, we used the fact that the twist matrix has no
dependence on the winding type coordinates so that all
$\tilde{\del}^{\mu}$ derivatives are zero and that they depend
only on the isometry coordinates so that $\del_{m} (U^{-1})^N_{\
B} =0$ (recall that $_{\mu} = (i, m)$, see Appendix
\ref{conventions}.)

When $U^{-1} = T_{{\rm NATD}}(\nu)$ in (\ref{NATDmatrix}) we have
\bea && (T_{{\rm NATD}})_i^{\ \ I}= 0, \ \ (T_{{\rm NATD}})_{iI} =
\delta_{iI}, \ \ \ (T_{{\rm NATD}})^{iI} = \delta^{iI}, \ \ \
(T_{{\rm NATD}})^{i}_{\ \ I} =\theta^i_{\ I}, \label{components}
\\ && (T_{{\rm NATD}})_m^{\ a} = \delta_m^{\
                    a}, \ (T_{{\rm NATD}})^m_{\ a} = \delta^m_{\ a}, \ (T_{{\rm NATD}})_m^{\ I}
                    = (T_{{\rm NATD}})^m_{\ I} = (T_{{\rm NATD}})_i^{\ a} = (T_{{\rm NATD}})^i_{\ a} = 0. \nonumber \eea where we have defined
\be \theta^i_{\ I} = \delta^{iJ}  C_{IJ}^{\ \ K} \nu_K,\ee  so
that the indices match. Plugging these in the formula we find that
the only non-vanishing components are \be \Omega^I_{\ JK} =
-C_{JK}^{\ \ I}, \ \ \ \ \Omega_{IJK} = C_{LI}^{\ \ H} C_{JK}^{\ \
L} \nu_{H}. \ee These give rise to the following fluxes \bea
f_{IJ}^{\ \ K} &=& \Omega_{IJ}^{\ \ K} + \Omega_{J \ \ \ I}^{\ \
K} + \Omega^K_{\ IJ}
= -C_{IJ}^{\ \ K} \nonumber \\
f_{IJK} &=& \Omega_{IJK} + \Omega_{JKI} + \Omega_{KIJ} =
\frac{1}{2} C_{L[I}^{\ \ \ H} C_{JK]}^{\ \ L} \nu_H = 0,\nonumber
\eea where the last equality follows from the Jacobi identity.

Now,  let us compute the fluxes associated with the geometric
twist matrix (\ref{geomtwist}) so that $U^{-1} = L(\theta)$ . In
this case the only non-vanishing flux is the geometric flux
$f_{IJ}^{\;\;\;\;\;K} = \Omega_{IJ}^{\ \ K} - \Omega_{JI}^{\ \ K}$
(since $\Omega^K_{\ IJ} = 0$):
\begin{equation}
f_{IJ}^{\;\;\;\;\;K}=-l^{i}_{\;I}\partial_{i}l^{j}_{\;J}l^{K}_{\;j}+l^{i}_{\;J}\partial_{i}l^{j}_{\;I}l^{K}_{\;j}
= -C_{IJ}^{\;\;\;\;\;K}.
\end{equation} This follows  from the fact that $\sigma^I = l^I_{\ i}
d\theta^i$ are left-invariant one-forms and as such  they satisfy
\be d\sigma^I = -\frac{1}{2} C_{IJ}^{\ \ K} \sigma^J \wedge
\sigma^K. \ee

The fluxes associated with $T_{{\rm NATD}}$ and $L$ are exactly
the same. This will be the key point in proving that NATD is a
solution generating transformation for DFT.


\subsection{Comparing the Field Equations of DFT and
GDFT}\label{dfteom}

In the previous subsection, we studied GDFT, which is obtained
from Scherk-Schwarz reduction of DFT. The Scherk-Schwarz anzats is
known to give rise to a consistent dimensional reduction, meaning
that any solution of the field equations of the resulting theory
can be uplifted to a solution of the higher dimensional field
equations \cite{SS}. In our case this implies that any solution of
the field equations of  GDFT  can be uplifted to a solution for
DFT. Conversely, the field equations of DFT will reduce to the
field equations of GDFT  and hence, given  a solution of DFT for
which the dependence of the fields on the doubled coordinates is
separated as in (\ref{anzats1},\ref{anzats2}) and (\ref{anzats3}),
the untwisted fields $\cH(x), F(x), d(x)$ will form a solution of
the GDFT equations, where the fluxes in the GDFT action is
determined by the twist matrix $U(Y)$.

This straightforward argument should be discussed in more detail,
mainly for two reasons. Firstly, the anzats in
(\ref{anzats1},\ref{anzats2}) and (\ref{anzats3}) is not exactly
the Scherk-Scwarz anzats which gives a consistent dimensional
reduction to GDFT, due to the difference in the anzats for RR
fields. The correct anzats would have been \be \label{gecici}
\chi(x,Y) = e^{-\sigma(Y)} S(Y) \chi(x)\ee which gives rise to a
deformation of the RR sector, as well. Although the field $\chi$
appears in the DFT action only through its field strength  $F =
e^{-B} \slashed{\del} \chi$, it has a bare appearance  in the
gauge transformation rules  and hence a consistent reduction
should involve an anzats for the field $\chi$.\footnote{Duality
twisted reduction of DFT with the anzats (\ref{gecici}) was
studied in \cite{Aybike}.} However, at the level of equations of
motion, this raises no problem since the field $\chi$  never
appears in the field equations  without a derivative. As a second
important point, the real duality group for DFT is $Spin^+(d,d)$
and hence only a twist matrix in this subgroup of $Spin(d,d)$ can
give a consistent reduction. This point is particularly important
for us, as the NATD matrix in (\ref{NATDmatrix}) is not in
$Spin^+(d,d)$. However, as discussed in \cite{dftRR}, although the
$Pin(d,d)$ transformations which are not in this subgroup are not
invariances of DFT,  they act as duality transformations. This is
also true at the level of field equations. In order to clarify
these points, we will discuss below the relationship between the
field equations of DFT and  of GDFT in more detail.

\subsubsection{Field Equations for the generalized dilaton field:}

The field equations obtained by varying the DFT action with
respect to the generalized dilaton field is \cite{HullZ4} \be
\label{dilatoneqn} \cR = 0, \ee where $\cR$ is as in
(\ref{ricci}). If we plug in (\ref{anzats1}) and (\ref{anzats3})
in (\ref{dilatoneqn}), we obtain \be \label{dilatoneqn2} \cR +
\cR_f = 0 \ee as was shown in \cite{Grana}. The form of $\cR_f$
was given in (\ref{deformedRicci2}). It can be easily shown that
this is the field equation obtained by varying (\ref{GDFT}) with
respect to the generalized dilaton field. Therefore, a set of DFT
fields whose dependence on the coordinates is separated as in
(\ref{anzats1}), (\ref{anzats2}) and (\ref{anzats3}) will satisfy
the generalized dilaton field equations of DFT if and only if the
untwisted fields $\cH(x), d(x)$ satisfy the generalized dilaton
equations for the GDFT, where the fluxes $f_{ABC}$ which determine
the deformation are determined by the twist matrix $U$.


\subsubsection{Field Equations for the spinor field $\chi$:} The field
equation for the spinor field $\chi$ is \cite{dftRR} \be
\slashed{\del}(\cK \ \slashed{\del}
   \chi) = 0, \ee which is to be supplemented with the duality constraint \be
\slashed{\del} \chi = -\cK \slashed{\del} \chi. \ee In terms of
the field $F = e^{-B} \slashed{\del}\chi$ the equation and the
duality constraint becomes: \bea \label{Feqn} \slashed{\del}(\cK \
e^{B} F) = 0 \\ \label{Feqn1} F = - e^{-B} \cK  e^{B} F. \eea
Imposing the duality constraint on the field equation we  get \be
\label{eomchi} \slashed{\del} (e^{B} F) = 0. \ee Let us plug in
(\ref{anzats1}) and (\ref{anzats2}) in (\ref{Feqn1}). We
immediately see that the duality constraint is satisfied by
$\cH(x,Y), F(x, Y)$ if and only if the same duality constraint is
satisfied by $F(x), \cH(x)$: \be \label{dualityL} F(x) = -
e^{-B(x)} \cK(x)  e^{B(x)} F(x). \ee On the other hand, plugging
(\ref{anzats1}) and (\ref{anzats2}) into the field equation
(\ref{Feqn}) we get \be \label{yeni1}
 \slashed{\del} \left(e^{-\sigma(Y)} S(Y) \cK(x) e^{B(x)}
F(x) \right) = 0.\ee We plug the duality constraint
(\ref{dualityL}) in (\ref{yeni1}) to get (recall that
$\slashed{\del} = \Gamma^M \del_M$): \be \label{Feqn2}
e^{-\sigma(Y)} S(Y) \left\{S^{-1}(Y)\Gamma^M \del_M (S(Y) -
\sigma(Y)) + \slashed{\del}\right\}\left(e^{B(x)} F(x)\right) =
0.\ee Now we use the following facts \cite{Aybike}: \footnote{We
proved the identity (\ref{isom}) in \cite{Aybike} for $S \in
Spin^+(d,d)$. It can be easily shown that it also holds for
elements of $ S \in Spin(d,d)$ of the form $S = C S^+$ and $S =
S^+ C$, where $S^+ \in Spin^+(d,d)$.} \be \label{esas} S^{-1}
\Gamma^M S = (U^{-1})^M_{\ A} \Gamma^A \ee \bea \Gamma^A \
(U^{-1})^M_{ \ \ A} S^{-1}
\partial_M \ S &=& \frac{1}{4} \Omega_{ABC} \Gamma^A \ \Gamma^B \
\Gamma^C \nonumber \\
& = & \frac{1}{12} f_{ABC} \Gamma^A \ \Gamma^B \ \Gamma^C
-\frac{1}{2} f_B \Gamma^B,  \label{isom} \eea where $U =
\rho(S^{-1})$.  Using these one can show easily that the equation
(\ref{Feqn2}) is equivalent to \be \label{Feqnesas}
\slashed{\nabla}\left(e^{B(x)} F(x) \right) = 0, \ee where the
Dirac operator $\slashed{\nabla}$ is defined as (see
\cite{Aybike}) \be \label{Dirac} \slashed{\nabla} \equiv
\slashed{\del} + \frac{1}{12} f_{ABC} \Gamma^A \ \Gamma^B \
\Gamma^C -\frac{1}{2} \eta_B \Gamma^B. \ee As a result, we
conclude that the fields (\ref{anzats1}) and (\ref{anzats2}) form
a solution for the  field equation (\ref{eomchi}) if and only if
the untwisted fields $F(x), \cH(x)$ satisfy
(\ref{Feqnesas}).\footnote{Note that (\ref{Feqnesas}), which is
equivalent to $ \slashed{\nabla}\left(\slashed{\del} \chi(x)
\right) = 0 $ is not the field equation obtained from varying
(\ref{GDFT}) with respect to the spinor field $\chi$, which would
have given $\slashed{\del}\left(\slashed{\del} \chi(x) \right) =
0$. It is not the  field equation obtained from varying the GDFT
action of the RR sector obtained in \cite{Aybike} through a
duality twisted ansazt on $\chi$ (rather than $F$) either, which
would have yielded $\slashed{\nabla}\left(\slashed{\nabla} \chi(x)
\right) = 0$. Note that both of these equations are satisfied
automatically due to nilpotency of $\slashed{\del}$ and
$\slashed{\nabla}$.}


\subsubsection{Field Equations for the generalized metric
$\cH_{MN}$:}\label{eomH}

The field equations obtained from varying the DFT action with
respect to the generalized metric $\cH_{MN}$ is \cite{HullZ4},
\cite{dftRR}: \be \label{genmetriceqn} R_{MN} + e^{-2\phi}
\Xi_{MN} = 0, \ee where
\begin{eqnarray} \label{RReom} \Xi^{MN} &=& \frac{1}{16}
\cH^{(M}_{\ \ P} \langle \slashed{\del}\chi, \Gamma^{N)P} \cK
\slashed{\del}\chi \rangle = \frac{1}{16} \cH^{(M}_{\ \ P} \langle
e^{B} F, \Gamma^{N)P} \cK e^{B} F \rangle \\ \nonumber &=& -
\frac{1}{16} \cH^{(M}_{\ \ P} \langle  F, e^{-B} \Gamma^{N)P}
e^{B} F \rangle.
\end{eqnarray} The first term in (\ref{genmetriceqn}) comes from the variation of the GDFT action of the NS-NS sector, and the variation of
the GDFT action of the RR sector gives the second term.   In
passing to the second line in (\ref{RReom}), we used the
invariance property of Mukai pairing under
$Spin^+(d,d)$\footnote{The Mukai pairing satisfies \be \langle S.
\phi_1, S. \phi_2 \rangle = \pm \langle \phi_1, \phi_2 \rangle, \
\ S \in Spin^{\pm}(d,d), \ee} (which $e^{-B}$ is an element of),
and we also imposed the duality constraint $\slashed{\del}\chi =
-\cK \slashed{\del}\chi$. Here, $\Gamma^{MN}$ is defined as
$\Gamma^{PQ} \equiv \frac{1}{2} [\Gamma^P , \Gamma^Q ]$. Let us
plug in the set of fields in (\ref{anzats1}, \ref{anzats2},
\ref{anzats3}) into these equations. Consider first the following
expression: \be \label{genmetriceqn1} (U^{-1})^M_{\ A}
R_{MN}[\cH(x, Y), d(x,Y)] (U^{-1})^N_{\ B}. \ee We emphasize again
that $ R_{MN}[\cH(x, Y), d(x,Y)]$ is obtained by varying $e^{-2d}
\cR$ with respect to $\cH^{MN}$ and then plugging in $\cH(x,Y)$.
Now compare the expression in (\ref{genmetriceqn1}) with the
variation of the GDFT action of the NS-NS sector (which is
obtained by plugging in $\cH(x,Y)$ in $\cR$ first) with respect to
$\cH^{AB}$. Comparing term by term, one sees that the two give the
same result. Then, we have \be e^{2d}(U^{-1})^M_{\ A}
R_{MN}[\cH(x, Y), d(x,Y)](U^{-1})^N_{\ B}  =
\frac{\delta(e^{2d}(\cR + \cR_f))}{\delta \cH^{AB}}. \ee So, if we
define \be \displaystyle e^{2d} R_f^{AB} = \frac{\delta (e^{2d}
\cR_f)}{\delta \cH_{AB}},\ee then we have \be \label{result1}
R^{MN}[\cH(x, Y), d(x,Y)] = (U^{-1})^M_{\ A}\left(R^{AB}[\cH(x),
d(x)] + R_f^{AB}[\cH(x), d(x)]\right)(U^{-1})^N_{\ B}. \ee Now, we
plug in $\Xi^{MN}$ the fields $\cH, \cK, F, B$, whose dependence
on the coordinates $(x,Y)$ is separated as  in (\ref{anzats1},
\ref{anzats2}, \ref{anzats3}). If we use the invariance property
of the Mukai pairing  and the following identity \be S^{-1}
\Gamma^{MN} S = \Gamma^{AB} (U^{-1})^M_{\ A} (U^{-1})^N_{\ B},\ee
along with \be \cH^M_{\ P}(x, Y) = (U^{-1})^M_{\ A} U^B_{\ P}
\cH^A_{\ B}, \ee we obtain
\begin{eqnarray} \label{eom} &&- \frac{1}{16} (U^{-1})^{M}_{\
A}(Y) (U^{-1})^{N}_{\ B}(Y) \cH^{(A}_{\ \ C}(x) \langle S(Y)
e^{B(x)} F(x), S(Y) \Gamma^{B)C} e^{B(x)} F(x) \rangle
\\ \nonumber &=& \mp \frac{1}{16}
(U^{-1})^{M}_{\ A}(Y)  \cH^{(A}_{\ \ C}(x) \langle e^{B(x)} F(x),
\Gamma^{B)C}  e^{B(x)} F(x) \rangle (U^{-1})^{N}_{\ B}(Y),\ \ S
\in Spin^{\pm}(10,10).
\end{eqnarray}

Therefore, we have found that \be e^{-2d(x,Y)}
\Xi^{MN}[\cH(x,Y),F(x,Y),d(x,Y)]= (U^{-1})^{M}_{\ A}(Y) e^{-2d(x)}
\ \Xi^{AB}[\cH(x), F(x), d(x)] \ (U^{-1})^{N}_{\ B}(Y), \ee if the
twist matrix $S(Y)$ is in $Spin^+(d,d)$.

Now consider the case when $S$ is not in $Spin^+(d,d)$. We assume
that it is of the form $S(Y) = S_1(Y) C$, where $S_1 \in
Spin^+(d,d)$ and $C$ is the charge conjugation element satisfying
$\rho(C) = J$. (This is the case for the twist matrices that
determines our NATD fields. Recall that $S_{{\rm NATD}}(\nu) =
S_{\beta}(\nu) C$ and $ S_{\beta} \in Spin^+(d,d)$.) In this case
only the $S_1(Y)$ factor can be dropped in passing from the first
line to the second line in (\ref{eom}) and we end up with \bea &
&- \frac{1}{16} (U^{-1})^{M}_{\ A}(Y) (U^{-1})^{N}_{\ B}(Y)
\cH^{(A}_{\ \ D}(x) \langle C e^{B(x)} F(x),  C \Gamma^{B)D}
e^{B(x)} F(x)
\rangle \\
&=& \label{sigmaeom} - \frac{1}{16} (U_1^{-1})^{M}_{\ A}(Y)
(U_1^{-1})^{N}_{\ B}(Y) \bar{\cH}^{(A}_{\ \ D}(x) \langle C
e^{B(x)} F(x), \Gamma^{B)D} C e^{B(x)} F(x) \rangle \eea where
$\rho(S_1) = U_1$ and $\bar{\cH} = J^T \cH J$ is the dual
generalized metric we defined in (\ref{dual1}). Note that in
writing the second line above we used \be \label{dualvar}
(U^{-1})^T \cH U^{-1} = ((U_1 J)^{-1})^T \cH (U_1 J)^{-1} =
(U_1^{-1})^T J^T\cH J U_1^{-1} = (U_1^{-1})^T \bar{\cH} U_1^{-1}.
\ee Also recalling the definition of the dual spinor field $
\bar{F}$ in (\ref{dual2}) we see that (\ref{sigmaeom}) above can
be written in terms of the dual fields and we have: \be
\Xi^{MN}[\cH(x,Y),F(x,Y),d(x,Y)] = (U_1^{-1})^{M}_{\ A}(Y)
\Xi^{AB}[\bar{\cH}(x), \bar{F}(x), d(x)] (U_1^{-1})^{N}_{\
B}(Y).\ee


Let us now try and write the $R_{MN}$ part of the field equation
in terms of the dual generalized metric $\bar{\cH}$, as well. For
this we need to observe that this piece of the field equation is
$O(d,d)$ covariant: \bea \label{oddcov3} R^{AB}[\cH(x)]
&=& R^{AB}[J \bar{\cH} J] = J R^{AB}[\bar{\cH}(x), d] J \\
\label{oddcov4} R_f^{AB}[\cH(x), d] &=& R_f^{AB}[J \bar{\cH}(x) J
d] = J R_{\bar{f}}^{AB}[\bar{\cH}(x), d] J \eea where $\bar{f}$ is
the dual flux we defined in (\ref{dualflux}). These follow
directly from the $O(d,d)$ covariance of $\cR$ and $\cR_f$, see
(\ref{oddcov2}). Recall that $J$ is obtained by embedding the
$O(d,d)$ matrix $J_d$  in $O(10,10)$ as in (\ref{embedding}).
Hence,  it acts non-trivially only  on the isometry directions and
acts on the partial derivatives with respect to  $x$ coordinates
as an identity transformation. Therefore, we have \be
\label{notwist} J^M_{\ \ A} \del_M g(x)  = \del_A g(x), \ee where
$g(x)$ denotes any of the untwisted  fields $\cH(x), F(x), d(x)$
or $\ciftS(x)$. So, we have $\bar{\del} = \del$ in
(\ref{oddcov2}). As a result, using (\ref{dualvar}), we can
rewrite (\ref{result1}) as: \bea R^{MN}[\cH(x, Y), d(x,Y)]
&=&(U^{-1})^M_{\ A} \left(R^{AB}[\cH(x), d(x)] + R_f^{AB}[\cH(x), d(x)]\right) (U^{-1})^N_{\ B} \nonumber  \\
& =& (U_1^{-1})^M_{\ A} R^{AB}[\bar{\cH}(x), d(x)] +
R_{\bar{f}}^{AB}[\bar{\cH}(x), d(x)] (U_1^{-1})^N_{\ B}. \nonumber
\eea

To recap, we have obtained the following: For $S \in Spin^+(d,d)$
with $\rho(S) = U^{-1}$, equation (\ref{genmetriceqn}) is
satisfied by the fields $\cH(x, Y), \ciftS(x, Y), F(x, Y)$ and
$d(x, Y)$ if and only if the untwisted fields satisfy the
following GDFT equation: \be \label{ek2} R^{AB}[\cH(x),d(x)] +
R_f^{AB}[\cH(x),d(x)] +e^{-2d(x)}\Xi^{AB}[\cH(x),F(x),d(x)] = 0.
\ee Here the fluxes $f$ in $R_f$ are produced by the twist matrix
$U$.

On the other hand, if $S = S_1 C$ with $S_1 \in Spin^+(d,d)$ and
$\rho(S_1) = U_1^{-1}$ so that $U =J U_1 $ we can make the
following statement. The twisted fields $\cH(x, Y), \ciftS(x, Y),
F(x, Y)$ and $d(x, Y)$ satisfy equation (\ref{genmetriceqn}) if
and only if the untwisted \emph{dual} fields $\bar{\cH}(x),
\bar{F}(x)$ and $d(x)$ satisfy \be \label{dualeqn}
R^{AB}[\bar{\cH}(x),d(x)] +
R_{\bar{f}}^{AB}[\bar{\cH}(x),d(x)]+e^{-2d(x)}\Xi^{AB}[\bar{\cH}(x),\bar{F}(x),d(x)]
= 0. \ee Here, the fluxes $\bar{f}$ in $R_{\bar{f}}$ are fluxes
\emph{dual} to $f$, and the fluxes $f$  are produced by the twist
matrix $U$.

\subsection{NATD Fields as a Solution of DFT in the Supergravity Frame}

We are now ready to prove our claim that NATD is a solution
generating transformation for DFT, that is, the  fields
(\ref{NATDH},\ref{NATDS}, \ref{NATDF}) corresponding to the NATD
background solve DFT equations. As we discussed before, this
immediately proves that the NATD fields form a solution of Type II
supergravity, if we identify $(x, \nu)$  with standard space-time
coordinates. This is because in the frame $\tilde{\del}^{\mu} = 0$
the DFT equations will reduce to Type IIA or Type IIB equations
depending on the fixed chirality of $\chi$\footnote{We still
assume that the duality group is unimodular. If not, the dilaton
field is forced to have a linear dependence on winding type
coordinates taking the NATD background out of the supergravity
frame. We will discuss this in section \ref{gse}}.

In the previous section, we saw that the fields $\cH(x,Y), F(x,Y),
\ciftS(x,Y)$ and $d(x,Y)$ in (\ref{anzats1}-\ref{anzats3}) satisfy
the field equations of DFT if and only if the untwisted  fields
$\cH(x), F(x), \ciftS(x)$ and $d(x)$ satisfy the field equations
of the GDFT determined by the fluxes associated with the twist
matrix $U$. This implies the following: Suppose that we know the
fields $\cH(x,Y), F(x,Y), \ciftS(x,Y)$ and $d(x,Y)$ satisfy the
field equations of DFT. Then, the untwisted fields satisfy the
field equations of GDFT determined by the fluxes associated with
$U$ and $S$. Now, consider another set of fields
$\tilde{\cH}(x,Z), \tilde{F}(x,Z)$, $\tilde{d}(x,Z)$ obtained by
twisting the same fields $\cH(x), F(x), d(x)$ by the twist
matrices $\tilde{U}(Z)$ and $\tilde{S}(Z)$, where $\tilde{U}$ is
also in $SO^+(d,d)$. Suppose also that the fluxes generated by
$\tilde{U}(Z)$ and $\tilde{S}(Z)$ are the same as the fluxes
generated by $U(Y)$ and $S(Y)$. Since we already know that the
untwisted fields satisfy the field equations of GDFT determined by
these fluxes, we immediately conclude that the twisted fields
$\tilde{\cH}(x,Z), \tilde{F}(x,Z)$, $\tilde{d}(x,Z)$  satisfy the
field equations of DFT, as well. If the NATD matrix
(\ref{NATDmatrix}) were in $SO^+(10,10)$, this argument would
immediately imply that the fields (\ref{NATDH}-\ref{NATDF}) formed
a solution of the DFT equations (\ref{dilatoneqn}), (\ref{eomchi})
and (\ref{genmetriceqn}), since we already know that the untwisted
fields $\cH(x), d(x), \ciftS(x)$ and $F(x)$ satisfy the GDFT
equations (\ref{dilatoneqn2}), (\ref{Feqnesas}) and (\ref{ek2}).
This is known because the fields $\cH(x,  \theta), \ciftS(x,
\theta)$ and $ F(x, \theta)$ in (\ref{separatedH}),
(\ref{separatedS}) and (\ref{separatedF}) form a solution of the
DFT equations (\ref{dilatoneqn}), (\ref{eomchi}) and
(\ref{genmetriceqn}) by construction, and the twist matrix
$L(\theta)$ in (\ref{geomtwist}) generates the same fluxes as the
NATD matrix (\ref{NATDmatrix}) does. However, the NATD matrix
$T_{{\rm NATD}}$ is not in $SO^+(10,10)$. Even in this case, our
argument above still holds when we compare the DFT and GDFT
equations (\ref{dilatoneqn})-(\ref{dilatoneqn2}) coming from the
variation with respect to the generalized dilaton field $d$ and
the equations (\ref{eomchi})-(\ref{Feqnesas}) coming from the
variation with respect to the spinor field $\chi$, since these
equations are not just $SO^+(d,d)$ covariant; they are covariant
under the full duality group $O(d,d)$. So, the only issue we
should discuss is how we compare  equations (\ref{ek2}) and
(\ref{dualeqn}).

In order to understand this, we look at a generic case in which
$U$ is in $SO^+(d,d)$, and $ \tilde{U} $ is not. We saw that
comparing the generalized metric field equations of DFT and the
GDFT is subtle due to the fact that the DFT of the RR sector of
Type II strings is invariant only under the subgroup $Spin^+(d,d)$
and $Pin(d,d)$ transformations that are not in this subgroup must
be viewed as dualities and not invariances. In analyzing this
case, we found it useful to define the following dual fields, as
in \cite{dftRR}, which we rewrite here for convenience: \be
\label{dual} \bar{\cH} = J \cH J, \ \ \bar{F} = e^{-\bar{B}} C
e^{B} F. \ee Recall that $\bar{F} = e^{-\bar{B}} \slashed{\del}
\bar{\chi}$, where  $ \bar{\chi} = C \chi.$ It is possible to
formulate the DFT action in terms of these dual fields. In fact,
it was shown in \cite{dftRR} that the DFT action takes the same
form in terms of these  dual fields as the action
(\ref{DFTaction}), provided that we also transform the
 partial derivatives as
$\del_i \leftrightarrow \tilde{\del}^i, \ i=1, \cdots, d$. We will
call this action the \emph{dual DFT action}. \footnote{In fact,
the DFT action of the RR sector picks up an overall minus sign but
so does the duality condition. Hence, when we plug in the duality
condition into the field equations, there is no overall minus sign
and the form of the field equations are exactly the same both in
terms of the original and the dual fields and coordinates.} If the
chirality of the spinor field $\chi$ is fixed in such a way that
the DFT action  reduces to the action of Type IIA/IIB theory in
the supergravity frame $\tilde{\del}^i = 0$, the dual DFT action
reduces to the action of Type IIB/IIA theory in the frame $\del_i
= 0$,  \cite{HullZ4, dftRR}. This is when  $d$ is odd. If $d$ is
even, the chirality of the dual spinor field remains the same, and
the dual action reduces to the same Type II action in the frame
$\del_i = 0$.\footnote{If the time direction is also dualized, the
resulting theory is Type IIA$^{\star}$ or Type IIB$^{\star}$
depending on the chirality, see \cite{dftRR}.}

Consider  a set of fields, which form a solution for the DFT field
equations in a certain frame\footnote{For now, we keep the
discussion general, but our ultimate goal is to apply the
discussion we have here to the fields (\ref{separatedH}),
(\ref{separatedS}) and (\ref{separatedF}).}. Then, the dual fields
will satisfy the equations arising from the dual DFT action for
the dual fields $\bar{\cH}$ and $\bar{F}$. We emphasize again that
these equations have exactly the same form as the equations for
the original fields, except that the standard derivatives along
the directions on which $J_d$ acts have been replaced by the
winding type derivatives and vice versa.\footnote{Let us clarify a
point that is potentially  confusing. When the frame in which the
fields satisfy the DFT equations is the supergravity frame (that
is, the fields have no dependence on dual coordinates
$\tilde{x}$), they also form a solution of Type IIA(/IIB)
supergravity. Since the dual fields will not belong to the frame
$\del_i = 0$ in general, they do not necessarily form a solution
of Type IIB(/IIA) supergravity. Nevertheless, they are a solution
of the field equations of the dual DFT action, and that is all the
information we need. In the special case when the isometry group
is Abelian, one can pick up coordinates with respect to which the
twisted fields will have no dependence on the coordinates $x^i, \
i=1, \cdots, d$ either, so the dual fields will belong to the
frame $\del_i =0$. Being a solution of the dual DFT equations,
they will hence form a solution of Type IIB(/IIA) supergravity.
This is what happens in Abelian T-duality.} If the dependence of
the fields forming the DFT solution on the coordinates $(x, Y)$ is
separated as in (\ref{anzats1},\ref{anzats2},\ref{anzats3}), then
the dependence of the dual fields on these coordinates  is also
separated in the following way: \be \label{danzats1} \bar{\cH}(x,
Y) = (\hat{U}^{-1})^T(Y) \bar{\cH}(x)\hat{U}^{-1}(Y), \ \
\bar{\cK}(x, Y) = \hat{S}(Y) \bar{\cK}(x) \hat{S}^{-1}(Y) \ee
\begin{equation} \label{danzats2}  \bar{F}(x, Y) = e^{-\sigma(Y)}e^{-\bar{B}(x,Y)} \hat{S}(Y)
e^{\bar{B}(x)}\bar{F}(x),
\end{equation} \be \label{danzats3} \bar{d}(x, Y) = \bar{d}(x) + \sigma(Y), \ee where $\hat{S} = C S C^{-1}, \ \rho(\hat{S}) = \hat{U}^{-1} = J U^{-1} J, $
and $\bar{F}(x,Y) = e^{-\bar{B}(x,Y)}C \slashed{\del}\chi(x,Y)$
and $\bar{d} = d$. Consider the field equation arising from
varying the dual DFT action with respect to the dual generalized
metric field and assume that it is satisfied by the dual fields
$\bar{\cH}(x, Y), \bar{F}(x, Y)$ and $d(x)$ in
(\ref{danzats1}-\ref{danzats3}). As emphasized above, this
equation is exactly of the same form as the generalized metric
field equation (\ref{genmetriceqn}) (albeit with $\del_i
\leftrightarrow \tilde{\del}^i$), which means that we can apply
the arguments in section \ref{eomH} directly. So, if $S$ is in
$Spin^+(d,d)$, so that $\hat{S} = C S C^{-1} \in Spin^+(d,d)$), we
find that the twisted dual DFT fields satisfy \be
\label{genmetriceqn2} R_{MN}[\bar{\cH}(x, Y), d(x)] + e^{-2\phi}
\Xi_{MN}[\bar{\cH}(x, Y), \bar{F}(x, Y), d(x)] = 0 \ee if and only
the untwisted dual DFT fields satisfy the following equation \be
\label{dualeqn2} (\hat{U}^{-1})^{M}_{\ A}(Y) \left((R +
R_{\hat{f}})^{AB}[\bar{\cH}(x), \bar{F}(x), d(x)]+ e^{-2d(x)}
\Xi^{AB}[\bar{\cH}(x), \bar{F}(x), d(x)]\right)
(\hat{U}^{-1})^{N}_{\ B}(Y) = 0. \ee  Since all the $\del_i$
derivatives in (\ref{genmetriceqn2}) has been swapped with the
winding type derivatives $\tilde{\del}^i$,   in calculating the
fluxes $\hat{f}$ with the formula (\ref{structure}) (with $U =
\hat{U}$), one should replace $\del_i \leftrightarrow
\tilde{\del}^i$. Now remember our discussion in section
(\ref{fluxes}). From (\ref{omegatilde}) we see that  the fluxes
$\hat{f}$ are produced by the matrix $\breve{U} = \hat{U} J$.
Since $\hat{U} = J U J$ we see that the fluxes $\hat{f}$ are the
same fluxes as those produced by the twist matrix $\breve{U} =
\hat{U} J = J U$, since $J^2 = Id$. But, according to
(\ref{omegabar}) this is just the dual flux $\bar{f}$ to the flux
$f$ produced by the twist matrix $U$, that is, $\hat{f} =
\bar{f}$. As a result, the equation (\ref{dualeqn2}) is equivalent
to the equation (\ref{dualeqn}). This gives us the result that we
want: the fact that the fields $\cH(x, \theta), F(x, \theta), d(x)
$ in (\ref{separatedH}, \ref{separatedF}) satisfy the DFT equation
(\ref{genmetriceqn}) implies that the dual  fields $\bar{\cH}(x,
\theta), \bar{F}(x, \theta), d(x)$ satisfy the dual DFT equation
(\ref{genmetriceqn2}). As a result, the
 dual untwisted fields $\bar{\cH}(x), \bar{F}(x), d(x)$ satisfy the
GDFT equation (\ref{dualeqn}), which then implies that the NATD
fields $\cH(x, \nu), F(x, \nu), d(x)$ in (\ref{NATDH}-\ref{NATDF})
satisfy the DFT equation (\ref{genmetriceqn}), as desired.

\subsection{Non-Unimodular Case: Generalized Supergravity
Equations}\label{gse}

So far, we have assumed that the isometry group $G$ is unimodular
so that the structure constants $C_{IJ}^{\ \ K}$ are traceless.
 When this assumption is
relaxed, it is known that the resulting NATD fields form a
solution of the GSE, which have recently been introduced in
\cite{Arutyunov:2015mqj}, \cite{Wulff:2016tju}. Let us see see how
this situation fits within the framework of DFT.

For simplicity, we assume that the structure constants of the Lie
algebra of $G$ have only trace components. Then, the only
non-vanishing components of the flux associated with the twist
matrix $L$ in (\ref{separatedH}) will be $f_I, \ I=1, \cdots, d$.
This contributes to $\eta_I$, whose definition is given in
(\ref{structure}). However, it is well-known that the GDFT action
with non-vanishing $\eta_A$ is not consistent \cite{Grana,
Geissbuhler}). Therefore, the $f_I$ part in (\ref{structure})
should be compensated by a non-trivial dilaton anzats. A similar
situation was also considered \cite{Catal-Ozer:2017ycb}. Rewriting
(\ref{structure}) in components, we see that we need to have \bea
\eta^I & = & f^I - (U^{-1})^{MI}
\del_M \sigma = 0, \\
\eta_I & = & f_I - (U^{-1})^M_{\ \ I} \del_M \sigma =0.  \eea This
implies that \bea (U^{-1})^{MI} \del_M \sigma &=& f^I = 0,
\\ (U^{-1})^M_{\ \ I} \del_M \sigma &=& f_I = {\rm
constant}. \eea When the twist matrix is equal to the NATD matrix
(\ref{NATDmatrix}), we can expand these equation by using
(\ref{components}) as:  \bea
\delta^{iI} \del_i \sigma   &=&  0 \\
\theta^i_{\ I} \del_i \sigma + \delta_{iI} \tilde{\del}^i \sigma
&=& {\rm constant}. \eea As a result, we obtain $$ \del_i \sigma =
0, \ \ \ \tilde{\del}^i \sigma = {\rm constant}.$$ In other words,
$\sigma$ is linear in the dual coordinates and  does not depend on
the standard coordinates. Then, the generalized dilaton field in
(\ref{NATDd}) is of the form: \be d(x, \tilde{\nu}) = d(x) + m_i
\tilde{\nu}^i, \ee where $m_i$ are constants.

Appearance of winding type coordinates in the transformed DFT
fields means that we are not in the supergravity frame anymore.
(Note that, due to the form of the anzats (\ref{NATDF}), the
spinor field $F$ also has a dependence on $\tilde{\nu}$). The
other DFT fields $\cH$ and $ \ciftS$ depend only on the space-time
coordinates. In the papers \cite{Yoshida1} and \cite{Yoshida2}, it
was shown that the equations of DFT  reduce to GSE in such a
frame. As a result, the fields in the target space of the NATD
model  form a solution of GSE, when $G$ is non-unimodular.


\section{Conclusions and Outlook}\label{conclusion}

In this paper, we studied NATD as a coordinate dependent $O(d,d)$
transformation. The dependence  on the coordinates  is determined
by the structure constants of the Lie algebra of the isometry
group $G$. Besides making calculations significantly easier, our
approach gives a natural embedding of NATD  in Double Field Theory
(DFT), a framework which provides an O(d,d) covariant formulation
for effective string actions \cite{Tseytlin1}-\cite{dftRR} by
introducing dual, winding type coordinates.  As a result of this
embedding, we managed to prove that the NATD fields (both in the
NS-NS and the RR sector) solve supergravity equations, when the
isometry algebra is unimodular. When the isometry algebra is
non-unimodular, we showed that the generalized dilaton field of
DFT is forced to have a linear dependence on the winding type
coordinates, which implies that the NATD fields solve GSE, in
agreement with the literature.

We believe that identifying the $O(d,d)$ matrix that generates the
NATD background is  important, as it should make it easier to
study some properties (such as supersymmetry and integrability) of
the NATD backgrounds and their CFT duals, as the relation to the
original background is more explicit. On the other hand, our
approach also makes it possible to explore the relation between
NATD and Yang-Baxter (YB) deformations in detail. Homogoneous YB
deformation of an integrable sigma model \cite{Kawaguchi:2014qwa}
is determined by the so called R-matrix, which forms a solution of
the classical Yang-Baxter equation. In the paper
\cite{Hoare:2016wsk}, it was conjectured that homogoneos YB models
can  be obtained by applying NATD to the original background, with
respect to an isometry group determined by the R-matrix. This
conjecture was proved in \cite{Borsato:2016pas} for the case of
Principal Chiral Models (PCM) and they extended their work to
homogenous YB deformations of more general sigma model than PCM's
in \cite{Borsato:2018idb}. Then, the results of our paper implies
that it should be possible to describe YB deformations also as
$O(d,d)$ transformations. This approach was also taken in the
papers \cite{Lust:2018jsx,Demulder:2018lmj},
\cite{Sakamoto:2017cpu}-\cite{Bakhmatov:2018bvp} (see also the
papers \cite{Araujo:2017jkb,Araujo:2017jap} for a related
approach). The methods we have developed in this paper should give
a  deeper insight on YB deformations and the relation between NATD
and YB deformations. We hope to come back to these issues in the
near future \cite{winp}.

\section*{Acknowledgments}
We would like to thank E.~O Colgain and N.~S.~Deger for comments
and discussions.  This work is partially supported by the Turkish
Council of Research and Technology (T\"{U}B\.{I}TAK) through the
ARDEB 1001 project with grant number 114F321, in conjunction with
the COST action MP1405 QSPACE. We also acknowledge the financial
support of Scientific Research Coordination Unit of Istanbul
Technical University (ITU BAP) under the project TGA-2018-41742.

\appendix

\section{Index Conventions}\label{conventions}
Our index conventions are as follows:

$M, N, \cdots $: Doubled coordinates; $ ^M = (_{\mu}, \ ^{\mu})$

$A, B, \cdots $: Doubled coordinates; $ ^A = (_{\alpha}, \
^{\alpha})$

$\mu = (i, m), \ \ \ \mu = 1, \cdots, 10; \ \ i=1, \cdots, d, \ \
d= {\rm dim} G$

 $\alpha = (I, a), \ \ \ \alpha = 1, \cdots, 10; \
\ I=1, \cdots, d$

\noindent  According to the embedding rules in (\ref{embedding}),
a twist matrix $T \in O(D,D,R)$, which only twists the $d$
isometry directions is of the following form: \be T^M_{\ A} =
\left(\begin{array}{cc}
                    T_{\mu}^{\ \alpha} & T_{\mu \alpha} \\
                    T^{\mu \alpha} & T^{\mu}_{\ \alpha}
                    \end{array}\right), \ee with $T_m^{\ a} = \delta_m^{\
                    a}, \ T^m_{\ a} = \delta^m_{\ a}, \ T_m^{\ I}
                    = T^m_{\ I} = T_i^{\ a} = T^i_{\ a} = 0$.

\end{document}